\documentclass[aps, prd, 12pt, eqsecnum, tightenlines, notitlepage, nofootinbib, floatfix]{revtex4-1}

\PassOptionsToPackage{unicode}{hyperref}

\usepackage{hyperref, graphicx, slashed, xcolor, amsmath}
\usepackage{feynmp}
\allowdisplaybreaks

\definecolor{red}{rgb}{1,0,0}

\newcommand{\beq}{\begin{equation}}
\newcommand{\eeq}{\end{equation}}
\newcommand{\bea}{\begin{eqnarray}}
\newcommand{\eea}{\end{eqnarray}}

\begin{document}

\normalsize
\vspace{-0.5cm}
\begin{flushright}
CALT-TH-2017-033\\
\vspace{1cm}
\end{flushright}


\title{Non-Gaussian Enhancements of Galactic Halo Correlations in Quasi-Single Field Inflation}

\author{Haipeng An$^{a,b}$, Michael McAneny$^a$, Alexander K. Ridgway$^a$ and Mark B. Wise$^a$}
\affiliation{$^a$ Walter Burke Institute for Theoretical Physics,
California Institute of Technology, Pasadena, CA 91125, USA}
\affiliation{$^b$ Department of Physics, Tsinghua University, Beijing 100084, China}

\begin{abstract}We consider a quasi-single field inflation model in which the inflaton interacts with a massive scalar field called the isocurvaton.  Due to the breaking of time translational invariance by the inflaton background, these interactions induce kinetic mixing between the inflaton and isocurvaton, which is parameterized by a constant $\mu$.  We derive analytic formulae for the curvature perturbation two-, three-, four-, five-, and six-point functions explicitly in terms of the external wave-vectors in the limit where $\mu$ and the mass of the isocurvaton $m$ are both much smaller than $H$.  In previous work, it has been noted that when $m/H$ and $\mu/H$ are small, the non-Gaussianities predicted by quasi-single field inflation give rise to long wavelength enhancements of the power spectrum for biased objects (\textit{e.g.}, galactic halos).  We review this calculation, and calculate the analogous enhanced contribution to the bispectrum of biased objects.  We determine the scale at which these enhanced terms are larger than the Gaussian piece.  We also identify the scaling of these enhanced parts to the $n$-point function of biased objects.
\end{abstract}
\maketitle

\section{Introduction}
The inflationary paradigm \cite{SKS} proposes an era in the very early universe during which the energy density is dominated by vacuum energy.  It explains why the universe is close to flat and the near isotropy of the cosmic microwave background radiation.  In addition, it has a simple quantum mechanical mechanism for generating energy density perturbations with wavelengths that are well outside the horizon in the early universe. The energy density perturbations resulting from inflation have an almost scale invariant Harrison-Zeldovich power spectrum. The simplest inflation models consist of a single scalar field $\phi$ called the inflaton.  The quantum fluctuations in the Goldstone mode $\pi$ associated with the breaking of time translation invariance by the inflaton~\cite{Cheung:2007st} source the energy density fluctuations. In the simplest of these single field inflationary models, the density perturbations are approximately Gaussian~\cite{Maldacena:2002vr}. 

Quasi-single field inflation \cite{Chen:2009zp} is a simple generalization of single field inflation that consists of a massive scalar field, the isocurvaton field $s$, that couples to the inflaton.  This coupling can give rise to significant non-Gaussianities in the correlators of $\pi$. The  Lagrange density in this model contains an unusual kinetic mixing of the form $\mu  {\dot \pi}  s$ that gives rise to a wealth of interesting phenomena.

In this paper, we study the effects of primordial non-Gaussianities on large scale structure. One complication that is not present for the microwave background radiation is that galaxies are biased objects. They do not trace the mass distribution but rather arise at special points, for example where the fluctuations in the mass density exceed some threshold. It was realized in~\cite{AGW} and \cite{Dalal:2007cu} that the power spectrum for biased objects can deviate significantly from Harrison-Zeldovich on large scales if the primordial mass density perturbations are non-Gaussian.  These effects have become known as scale-dependent bias and stochastic bias.  In \cite{Baumann:2012bc} these enhancements for the power spectrum of biased objects were systematically explored within the context of quasi-single field inflation.\footnote{We refer to these effects as enhancements even though for some range of wave-vectors and model parameters they can interfere destructively with the usual part arising from  Gaussian primordial density fluctuations.} Quantitative predictions for the power spectrum of galactic halos in quasi-single field inflation (and other models for non-Gaussian primordial fluctuations) were recently made in~\cite{Gleyzes:2016tdh}. Very recently the scale-dependent bias introduced by higher spin fields~\cite{Arkani-Hamed:2015bza} coupled to the inflaton has been explored~\cite{MoradinezhadDizgah:2017szk}.

In this paper we continue and extend the work of \cite{Baumann:2012bc} and compute the galactic halo power spectrum and bispectrum in quasi-single field inflation. The bispectrum for galaxies was computed for local non-Gaussianity in \cite{Tellarini:2015faa} and for equilateral non-Gaussianity \cite{Mizuno:2015qma}.  We make explicit numerical predictions by adopting the very simple model in which galaxies arise at points where the underlying energy density fluctuations (averaged over a volume) are above a threshold~\cite{Press:1973iz}.\footnote{Kaiser applied this model to explain the biasing of rich clusters of galaxies~\cite{Kaiser}.}  Also, we identify the scaling of the $n$-point function of the halo overdensity in quasi-single field inflation within this threshold model. 

The impact of the non-Gaussianities in quasi-single field inflation is largest when the kinetic mixing $\mu$ and the isocurvaton mass $m$ are small compared to the Hubble constant during inflation $H$. We derive new analytic methods to calculate the correlations of $\pi$ in this region of parameter space. These are applied to derive analytic expressions for the two-, three-, four-, five-, and six-point functions of $\pi$. We apply these results to derive explicit expressions for the galactic halo power spectrum and bispectrum.  The effects in the power spectrum and the bispectrum of galaxies due to primordial non-Gaussianities can become pronounced at the scale $q \simeq 1/ (200 h^{-1}{\rm Mpc})$. In this work we neglect the time evolution of the galaxy distribution after galaxies form. Even though this is not a small effect, we do not expect that neglecting it will qualitatively impact our conclusions. Furthermore, the computations we perform of the higher correlations of $\pi$ will be useful for a more complete computation of the galaxy bispectrum.

In section II we outline the quasi-single field inflation model.  We discuss the power series expansion of the mode functions of the quantum fields $\pi$ and $s$ at small $|\tau|$, where $\tau$ is conformal time.  For small $\mu/H$ and $m/H$, a method is developed to determine the power series coefficients needed to compute the two-, three-, four-, five- and six-point correlations of the curvature perturbation $\zeta$.\footnote{$\pi$ and $\zeta$ are linearly related.}
In section III we compute the three-, four-, five- and six-point correlations of $\zeta$. The three- and four-point functions are computed for general wave-vectors, but the five- and six-point functions are only computed for the configurations of wave-vectors that are relevant to the long wavelength enhancements to the galactic halo bispectrum.
Section IV introduces the bias expansion and the points above threshold model for the galactic halo overdensity. The results from Section III are used to compute the halo power spectrum and bispectrum.  We also present the scaling of the $n$-point function of the halo overdensity in quasi-single field inflation. Concluding remarks are given in section V.

\section{The model and the mode functions}

We consider a quasi-single field inflation theory in which inflation is driven by a single scalar inflaton field $\phi$ and the inflaton is coupled to a single massive scalar isocurvaton field $s$.  The classical background field of the inflaton, $\phi_0(t)$, is time-dependent but we will impose conditions so that to leading order in slow-roll parameters, the background value of $s$ is zero.  We also impose a shift symmetry $\phi \rightarrow \phi + c$ and a $Z_{2}$ symmetry $\phi \rightarrow -\phi$ on the inflaton that is only broken by its potential. This implies that the isocurvaton field $s$ couples to derivatives of the inflaton. The lowest dimension operator coupling the inflaton to the isocurvaton is the dimension five operator, 
\begin{equation}
{\cal L}_{{\rm dim} ~5}= \frac{1}{\Lambda}{g^{\mu \nu}\partial_{\mu}\phi \partial_{\nu} \phi} s.
\end{equation}
We choose the gauge in which the inflaton is only a function of time, $\phi(x) = \phi_0(t)$. We expand the potential for $s$ in a power series about $s=0$, $V(s)=V's +V'' s^2/2 + V'''s^3/3! +...  $ and assume the tadpole in $s$  cancels, $(\dot \phi_0)^2/\Lambda -V'=0$. Since we work to leading order in slow-roll parameters, we can neglect ${\ddot{\phi}_0}$, making this cancellation possible.  To obtain long wavelength enhancements to the correlations of biased objects, we need $m$, the mass of $s$  ($m^2=V''$), to be less than the Hubble constant during inflation, $H$. We assume there is some inflaton potential (likely non-analytic in $\phi$) that gives values of the power spectrum tilt $n_S$ and the tensor to scalar ratio $r$ consistent with observations.

The Goldstone field $\pi(x)$, associated with time translational invariance breaking by the time dependence of $\phi_0$, gives rise to the curvature fluctuations.
In a de-Sitter background, the Lagrangian describing  $\pi(x)$ and $s(x)$ is
\begin{align}
\mathcal{L} = \mathcal{L}_{0} + \mathcal{L}_{{\rm int}}
\end{align}
where
\begin{equation}
\label{fft}
\mathcal{L}_{0}=\frac{1}{2(H\tau)^{2}}\left(\left(\partial_{\tau}\pi\right)^{2} - \nabla\pi\cdot\nabla\pi + (\partial_{\tau}s)^{2} - \frac{m^{2}}{(H\tau)^{2}}s^{2}-\nabla s\cdot\nabla s-\frac{2\mu}{H\tau}s\partial_{\tau}\pi\right)
\end{equation}
and
\begin{equation}
\label{intlagrange}
{\cal L}_{\rm int}  =  \frac{1}{(H\tau)^{4}}\left(\frac{(H\tau)^{2}}{\Lambda} \left((\partial_{\tau}\pi)^{2}- {\nabla { \pi}} \cdot  {\nabla \pi} \right)s - \frac{V'''}{3!} s^3 - \frac{V^{(4)}}{4!} s^{4} \ldots \right) .
\end{equation}
In eq.~(\ref{fft}) we have introduced
\begin{equation}\label{eq:mu}
 \mu=2{\dot { \phi}_0}/{{ \Lambda}}
 \end{equation}
and conformal time $\tau= -e^{-Ht}/H$.  We have rescaled $\pi$ by $\dot{\phi}_{0}$ (we take $\dot{\phi_0} >0$) to obtain a more standard normalization for the $\pi$ kinetic term.  We have also included the measure factor $\sqrt{-g}$ in the Lagrangian so that the action is equal to $\int d^3x  d\tau  {\cal L}$.
Note the unusual kinetic mixing term in (\ref{fft}) which is a result of the background inflaton field breaking Lorentz invariance. 

To compute correlation functions involving $\pi$ and $s$, we expand the quantum fields in terms of creation and annihilation operators. Since the fields $\pi$ and $s$ have kinetic mixing, they share a pair of creation and annihilation operators.  Introducing $\eta=k \tau$ we write,
\begin{equation}\label{eq:mode}
 \pi({\bf x},\tau)=\int {d^3 k \over (2 \pi)^3} \left( a^{(1)}({\bf k})  \pi_{k}^{(1)}(\eta) e^{i{\bf k}\cdot {\bf x}}+a^{(2)}({\bf k}) \pi_{k}^{(2)}(\eta)e^{i{\bf k} \cdot {\bf x}}  +{\rm h.c.}  \right)
\end{equation}
and
\begin{equation}\label{eq:mode2}
 s({\bf x},\tau)=\int {d^3 k \over (2 \pi)^3} \left( a^{(1)}({\bf k})  {s}_{k}^{(1)}(\eta) e^{i{\bf k}\cdot {\bf x}}+a^{(2)}({\bf k}) {s}_{k}^{(2)}(\eta)e^{i{\bf k} \cdot {\bf x}}  +{\rm h.c.}  \right)
 \end{equation}
By varying (\ref{fft}) we can obtain the equations of motion for the mode functions $\pi_{k}^{(i)}(\eta)$ and $s_{k}^{(i)}(\eta)$.  These are
\begin{equation}\label{eq:diff1}
\pi^{(i)\prime \prime}_k - \frac{2\pi^{(i) \prime}_k}{\eta} + \pi^{(i)}_k - \frac{\mu}{H} \left(  \frac{s^{(i)\prime}_k}{\eta} - \frac{3s^{(i)}_k}{\eta^2}\right) = 0 \ 
 \end{equation}
and
\begin{equation}\label{eq:diff2}
s^{(i)\prime \prime}_k-\frac{2s^{(i) \prime}_k}{\eta} + \left(1+\frac{m^2}{H^2\eta^2}\right)s^{(i)}_k + \frac{\mu}{H} \frac{\pi^{(i) \prime}_k}{\eta} = 0 \ ,
\end{equation}
 where  a `` $'$ '' indicates an $\eta$ derivative.  

\subsection{Power Series Solution}
As mentioned in the introduction it is difficult to solve equations (\ref{eq:diff1}) and (\ref{eq:diff2}) analytically for general $m$ and $\mu$.  Fortunately, in the small $m/H$ and $\mu/H$ regime we do not need the mode functions' full time-dependence to determine the leading behavior of the correlation functions of $\pi$.  Rather, we only need their small $-\eta$ behavior.\footnote{Conformal time $\eta$ satisfies $-\infty<\eta < 0$ with inflation ending at $\eta=0$.}  To determine this, we obtain a power series solution to (\ref{eq:diff1}) and (\ref{eq:diff2}). To begin, we rescale the mode functions
\begin{align}
\label{rescaled mode functions}
\pi^{(i)}_k (\eta) &= (H/k^{3/2}) \pi^{(i)} (\eta)\ \ \ \ \ \ \ \ s^{(i)}_k (\eta)= (H/k^{3/2}) s^{(i)} (\eta)
\end{align}
and then expand $\pi^{(i)}(\eta)$ and $s^{(i)}(\eta)$ as a power series in $-\eta$
\begin{align}
\label{series solution}
 \pi^i(\eta)=\sum_{n=0}^{\infty}a^{(i)}_{r,n}(-\eta)^{n+r}\ \ \ \ \ \ \ \ s^{(i)}(\eta) = \sum_{n=0}^{\infty}b^{(i)}_{r,n}(-\eta)^{n+r}.
\end{align}
By plugging (\ref{series solution}) into (\ref{eq:diff1}) and (\ref{eq:diff2}), we derive relations among the coefficients $a^{(i)}_{r,n}$ and $b^{(i)}_{r,n}$
\begin{align}
\label{coefficient equations}
&\left[a^{(i)}_{r,0}r-{\mu \over H} b^{(i)}_{r,0}\right](r-3)(-\eta)^{r-2} + \left[a^{(i)}_{r,1}(r+1) - {\mu \over H}b^{(i)}_{r,1}\right](r-2)(-\eta)^{r-1}\cr 
&\ \ \ \ + \sum_{n=0}^{\infty}\left[\left[a^{(i)}_{r,n+2}(n+r+2)-{\mu \over H} b^{(i)}_{r,n+2}\right](n+r-1) + a^{(i)}_{r,n}\right](-\eta)^{n+r} = 0\cr
&\left[\left[b^{(i)}_{r,0}(r-3)+{\mu \over H} a^{(i)}_{r,0}\right]r + b^{(i)}_{r,0}{m^{2}\over H^2}\right](-\eta)^{r-2} + \left[\left[b^{(i)}_{r,1}(r-2) + {\mu \over H} a^{(i)}_{r,1}\right](r+1) + b^{(i)}_{r,1}{m^{2}\over H^2}\right](-\eta)^{r-1}\cr 
&\ \ \ \ + \sum_{n=0}^{\infty}\left[\left[b^{(i)}_{r,n+2}(n+r-1)+{\mu \over H} a^{(i)}_{r,n+2}\right](n+r+2) + b^{(i)}_{r,n+2}{m^{2}\over H^2} + b^{(i)}_{r,n}\right](-\eta)^{n+r} = 0.
\end{align}
Since (\ref{coefficient equations}) is true for all $\eta<0$, the coefficient multiplying each power of $-\eta$ vanishes.  The constraints due to the coefficients multiplying $(-\eta)^{n+r}$ provide recursion relations relating the $n+2$ coefficients to the $n$ ones.  The constraints due to the coefficients multiplying $(-\eta)^{r-2}$ are
\begin{align}
\label{zero equations}
(a^{(i)}_{r,0}r - {\mu \over H} b^{(i)}_{r,0})(r-3) = 0,\ \ \ \ \ \left[b^{(i)}_{r,0}(r-3) + {\mu \over H} a^{(i)}_{r,0}\right]r + b^{(i)}_{r,0}{m^{2} \over H^2} = 0.
\end{align}
Equation (\ref{zero equations}) implies the only possible values of $r$ are 
\begin{align}
\label{r values}
r = 0, 3,\ \alpha_{-},\ \alpha_{+}
\end{align}
where 
\begin{align}
\label{alpha pm}
\alpha_{\pm} = 3/2 \pm \sqrt{9/4-\left(\mu/H\right)^{2} - \left(m/H\right)^{2}}.
\end{align}  
Note $\alpha_{-}$ and $\alpha_{+}$ approach 0 and 3 when $m$ and $\mu$ approach zero.  Then small $\mu/H$ and $m/H$ imply small $\alpha_{-}$.  Considering odd $n$ instead of even $n$ results in the same exact solution, so we take $a_{r,1}^{(i)}=b_{r,1}^{(i)}=0$ to eliminate this redundant solution.  

There are then four branches of the series solution (\ref{series solution}).  The leading power of each branch is $(-\eta)^{r}$ and the successive terms go like $(-\eta)^{r + 2 k}$ where $k$ is a positive integer.  The series solutions (\ref{series solution}) are a linear combination of each branch.  The small $-\eta$ behavior of $\pi^{(i)}$ and $s^{(i)}$ is then
\begin{align}
\label{small mode function behavior}
\pi^{(i)}(\eta) &= a^{(i)}_{0}+ a^{(i)}_{-}(-\eta)^{\alpha_{-}} + a^{(i)}_{0,2}(-\eta)^{2}+a^{(i)}_{-,2}(-\eta)^{\alpha_{-}+2}+ a^{(i)}_{+}(-\eta)^{\alpha_{+}} + a^{(i)}_{3}(-\eta)^{3}  + \dots\cr
s^{(i)}(\eta) &= b^{(i)}_{-}(-\eta)^{\alpha_{-}}+b^{(i)}_{0,2}(-\eta)^{2} + b^{(i)}_{-}(-\eta)^{\alpha_{-}+2} + b^{(i)}_{+}(-\eta)^{\alpha_{+}}+b^{(i)}_{3}(-\eta)^{3} + \dots
\end{align}
Note that we have used the notation $a_{\pm,n}^{(i)}\equiv a_{\alpha_\pm,n}^{(i)}$ and $b_{\pm,n}^{(i)}\equiv b_{\alpha_\pm,n}^{(i)}$, and we have also written the $n=0$ coefficients as $a_{r}^{(i)}$.  Moreover, $b_0^{(i)}=0$ due to ~(\ref{zero equations}).

As $-\eta \rightarrow 0$, $s^{(i)}(\eta)\rightarrow 0$ while $\pi^{(i)}(\eta) \rightarrow a^{(i)}_{0}$.  However, for $\alpha_{-} << 1$ the $(-\eta)^{\alpha_{-}}$ term will remain significant even for $-\eta << 1$ which means $\pi$ can undergo superhorizon evolution.  We can estimate the value of $\eta$ at which $\pi$ stops evolving using $\alpha_{-} \simeq \left(\mu^{2} + m^{2}\right)/(3H^{2})$ which is valid for small $\mu$ and $m$.  The $\pi$ modes then stop evolving at $-\eta \sim e^{-3H^2/\left(\mu^{2} + m^{2}\right)}$.  In this paper we only consider values of $m$ and $\mu$ such that the modes of interest stop evolving before the end of inflation. Then one does not need to consider the details of reheating to make predictions for the curvature perturbations.

Equation (\ref{zero equations}) can also be used to relate the $a^{(i)}$ and $b^{(i)}$ coefficients multiplying the leading $(-\eta)^{r}$ term of each branch
\begin{align}
\label{btoas}
b^{(i)}_{0} = 0,\ b^{(i)}_{-} = \frac{H a^{(i)}_{-} \alpha_{-}}{\mu},\ b^{(i)}_{+} = \frac{H a^{(i)}_{+} \alpha_{+}}{\mu}, b^{(i)}_{3} = \frac{-3 H\mu}{m^{2}}a^{(i)}_{3}.
\end{align}
A full solution to the mode equations is unnecessary.  We only need certain combinations of the power series coefficients to derive the leading (for small $m$ and $\mu$) behavior of the correlation functions of $\pi$ and $s$.  For example, the combinations $\sum\limits_{i}|a^{(i)}_{0}|^{2}$, $\sum\limits_{i}a^{(i)}_{0}b^{(i)*}_{-}$ and $\sum\limits_{i}|b^{(i)}_{-}|^{2}$ determine the two point functions $\left<\pi \pi\right>$,  $\left<\pi s\right>$ and $\left<s s\right>$ at late times.  

\subsection{Power Series Coefficients}
\label{commutation relations section and other stuff}
In this section, we outline the derivation of the combinations of power series coefficients that are needed to compute the correlation functions of $\pi$ when $m/H$ and $\mu/H$ are small.  We begin with the combination $\sum\limits_{i}|b^{(i)}_{-}|^{2}$, which can be obtained by matching to an effective field theory that reproduces the correct two point function of $s$ in the small $\eta$ limit.  It turns out that once we know $\sum\limits_{i}|b^{(i)}_{-}|^{2}$ we can determine $\sum\limits_{i}|a^{(i)}_{0}|^{2}$ and $\sum\limits_{i}a^{(i)}_{0}b^{(i)*}_{-}$ from the full theory. 

In the small $-\eta$ limit we can neglect the second term appearing in (\ref{fft}).  Then:
\begin{align}
\label{EFT2}
\mathcal{L}_0^{{\rm EFT}} = \frac{1}{2(H\tau)^{2}}\left(\left({\partial_{\tau} \pi}\right)^{2}+ (\partial_{\tau}s)^{2} - \frac{m^{2}}{(H\tau)^{2}}s^{2}-\nabla s\cdot\nabla s-\frac{2\mu}{H\tau}s{\partial_{\tau}\pi}\right)
\end{align}
The $\pi$ equation of motion gives
\begin{align}
\label{small tau pi}
\partial_{\tau}\pi = \frac{\mu}{H}\frac{s(\tau)}{\tau} 
\end{align}
where we have dropped a term proportional to $\tau^2$ in (\ref{small tau pi}). The solution of eq.~(\ref{small tau pi}) is
\begin{align}
\label{small tau pi in terms of s}
\pi(\tau) = c_{1} + \int\limits_{-\infty}^{\tau}\frac{\mu}{H}\frac{s(\tau^{\prime})}{\tau^{\prime}}d\tau^{\prime}
\end{align}
where $c_{1}$ is a constant operator.  As mentioned earlier, since (for small $\eta$) $s^{(i)}_{k}(\eta) \simeq b^{(i)}_{-}(-\eta)^{\alpha_{-}}$ and $\alpha_{-}$ is small, the mode functions $s^{(i)}_{k}$ remain nonzero even after the mode wave-vector has exited the horizon (i.e., when $|\eta|<1)$.  Due to the factor of $1/\tau$ in the integral in (\ref{small tau pi in terms of s}), the $\pi$ mode functions will undergo superhorizon growth and can become quite large if $m/H$ and $\mu/H$ are small.

We use eq.~(\ref{small tau pi in terms of s}) to express the field $\pi$ in terms of $s$. Integrating out $\pi$ using its equation of motion yields an effective Lagrangian for $s$:
\begin{align}
\label{EFT}
\mathcal{L}_0^{{\rm EFT}} = \frac{1}{2(H\tau)^{2}}\left((\partial_{\tau}s)^{2} - \frac{m^{2}+\mu^2}{(H\tau)^{2}}s^{2}-\nabla s\cdot\nabla s \right).
\end{align}
Since in this effective theory there is only one field $s$, it can be written in terms of a single mode function $s_k$ that satisfies the differential equation,
\begin{align}
\label{s mode equation}
s_k''(\eta) - \frac{2}{\eta}s_k'(\eta) + s_k(\eta) + \left(\frac{\mu^{2}}{H^{2}} + \frac{m^{2}}{H^{2}}\right)\frac{s_k(\eta)}{\eta^{2}} = 0.
\end{align}
The solution to (\ref{s mode equation}) that satisfies the asymptotic Bunch-Davies vacuum condition and is consistent with the canonical commutation relations is
\begin{align}
\label{s mode solution}
s_{k}(\eta) = H\sqrt{\frac{\pi}{4k^{3}}}(-\eta)^{3/2}H^{(1)}_{\nu}(\eta)
\end{align}
where $\nu = \sqrt{9/4 - (\mu/H)^{2} - (m/H)^{2}}$ and $H_\nu^{(1)}$ is a Hankel function of the first kind.  The small $-\eta$ limit of (\ref{s mode solution}) is 
\begin{align}
\label{small s mode function}
s_{k}(\eta) = H(-\eta)^{\alpha_{-}}\frac{i}{k^{3/2}}\frac{1}{\sqrt{2}}.
\end{align}
Using (\ref{small s mode function}), we can determine the small $-\eta$ limit of the two-point function of the Fourier transform of $s$. Denoting this Fourier transform by $s_{\bf k}$, we obtain
\begin{align}
\label{2pt:ss}
\left<s_{ \bf k}s_{\bf k'}\right>(\tau') = (2\pi)^{3}\delta^{3}\left({\bf k} + {\bf k}'\right)\frac{H^{2}}{2k^{3}}(-\eta)^{2\alpha_{-}}.
\end{align}
By matching the full theory prediction for $\left< s s \right>$ to (\ref{2pt:ss}) we find
\begin{align}
\label{b minus squared}
\sum\limits_{i}\left|b_{-}^{(i)}\right|^{2} = \frac{1}{2}.
\end{align}
Equation (\ref{small tau pi in terms of s}) can be used to determine the leading small $-\eta$ behavior of the $\pi$ mode functions in the full theory.  It gives
\begin{align}
\label{pi 0 in terms of s}
\pi^{(i)}(0) = c^{(i)}_{1} + \int\limits_{-\infty}^{0}\frac{\mu}{H}\frac{s^{(i)}(\eta^{\prime})}{\eta^{\prime}}d\eta^{\prime}.
\end{align}
From equation (\ref{small mode function behavior}) we see that the integrand in (\ref{pi 0 in terms of s}) goes like $(-\eta)^{-1 + \alpha_{-}}$ in the IR region of the integral, i.e. $-\eta < 1$.  For small $m/H$ and $\mu/H$, $\alpha_{-}$ is very small and the integral will receive a large contribution from the IR.  On the other hand, the contribution from the UV is small because the mode functions become oscillatory with smaller amplitude when $-\eta > 1$.  This means the integral is fixed by the integrand's IR behavior so that\footnote{We will use these same arguments when we evaluate the time integrals involved in the calculation of higher point correlators.}
\begin{align}
\label{eft pi zero}
\pi^{(i)}(0) \simeq c^{(i)}_{1} - \frac{\mu b_{-}^{(i)}}{H}\int\limits_{-1}^{0}(-\eta)^{-1 + \alpha_{-}}d\eta = c^{(i)}_{1} - \frac{\mu b_{-}^{(i)}}{H}\frac{1}{\alpha_{-}} = c^{(i)}_{1} - \frac{3\mu Hb_{-}^{(i)}}{\mu^{2} + m^{2}}.
\end{align}
In (\ref{eft pi zero}) we have used $\alpha_{-}^{-1} \simeq 3H^{2}/\left(\mu^{2} + m^{2}\right)$.  The corrections to (\ref{eft pi zero}) are suppressed by  powers of $\alpha_{-}$  and are unimportant when $m/H$ and $\mu/H$ are small.  The integral is insensitive to the exact value of the UV cutoff because $\alpha_{-}$ is small.

We can now compute the two-point function of the Fourier transform of $\pi$, which can be written as
\begin{align}
\label{2pt:pipi}
\left<\pi_{\bf k}(0)\pi_{\bf k'}(0)\right> \simeq (2\pi)^{3}\delta\left({\bf k} + {\bf k}'\right)\frac{H^{2}}{k^{3}} C_{2}(\mu, m).
\end{align} 
We determine $C_{2}(\mu, m)$ by taking the magnitude squared of (\ref{eft pi zero}):
\begin{align}
\label{C2 full exp}
C_{2}(\mu, m) \simeq \sum\limits_{i}\left[\left|c^{(i)}_{1}\right|^{2} + \frac{9 \mu^{2} H^{2}}{\left(\mu^{2} + m^{2}\right)}\left|b^{(i)}_{-}\right|^{2} - \frac{6\mu H}{\mu^{2} + m^{2}}{\rm Re}\left(c_{1}^{(i)}b_{-}^{(i)*}\right)\right].
\end{align}
In writing (\ref{C2 full exp}), we have only kept the terms that are most important for $m/H$ and $\mu/H$ small.  Now  $\left< \pi \pi  \right>$  is invariant under $s \rightarrow -s$.\footnote{If we treat $\mu$ as a perturbation then all of the corrections to $\left< \pi \pi  \right>$ involve even powers of the $s$ field.}  This implies the last term in the brackets of (\ref{C2 full exp}) has to vanish.  We can determine the first term by noting that the constant $c_{1}^{(i)}$ is $\mu$ independent.  This can be seen from the fact that it is a boundary condition fixed by the UV, thereby independent of the mixing factor $\mu$.  We can then fix the first term in (\ref{C2 full exp}) by demanding that $C_{2}(0,m) = 1/2$.  Finally, using (\ref{b minus squared}) we find that
\begin{align}
\label{leading C2}
C_{2}(\mu, m) \simeq \frac{1}{2} + \frac{9\mu^{2}H^{2}}{2\left(\mu^{2} + m^{2}\right)^{2}}.
\end{align}
Equation (\ref{leading C2}) gives the leading behavior of $C_{2}(\mu, m)$ in the limit of small $m/H$ and $\mu/H$.  We can determine the accuracy of (\ref{leading C2}) by extending the numerical techniques developed in \cite{Assassi:2013gxa} and \cite{An:2017hlx} to the region of small $m/H$ and $\mu/H$ and computing	 the power spectrum numerically.  This is done in appendix \ref{numerical checks appendix}.   

We now compute the leading $m$ and $\mu$ dependence of the curvature perturbation two-point function.  The curvature perturbation is related to the Goldstone field by
\begin{align}
\zeta_{\bf k} = -\frac{H}{\dot{\phi_0}}\pi_{\bf k}.
\end{align}
The curvature perturbation two-point is then
\begin{align}
\label{power spectrum}
\left<\zeta_{{\bf k}_{1}}\zeta_{{\bf k}_{2}}\right> = \left(\frac{H}{\dot{\phi}_{0}}\right)^{2}\left<\pi_{{\bf k}_{1}}\pi_{{\bf k}_{2}}\right> &= (2\pi)^{3}\delta({\bf k}_{1} + {\bf k}_{2}){\cal P}_{\zeta}(k)\cr 
&= (2\pi)^{3}\delta({\bf k}_{1} + {\bf k}_{2})\left(\frac{H^{2}}{\dot{\phi}_{0}}\right)^{2}\frac{1}{k^{3}} C_{2}(\mu,m).
\end{align}
Using (\ref{power spectrum}) we can express $\dot{\phi}_{0}$ in terms of $\mu$, $m$, and the measured value of the dimensionless power spectrum $\Delta_{\zeta}$ \cite{Ade:2015xua}:
\begin{align}
\Delta_{\zeta}^2 = 2.12 \times 10^{-9} = \frac{k^{3}}{2\pi^{2}}{\cal P}_{\zeta}(k) = \left(\frac{H^{2}}{\dot{\phi}_{0}}\right)^{2}\frac{1}{2\pi^{2}}C_{2}(\mu,m).
\end{align}
This implies
\begin{align}
\label{phi not dot}
\frac{\dot{\phi}_{0}}{H^{2}} = \sqrt{\frac{C_{2}(\mu,m)}{2\pi^{2}\Delta_{\zeta}^2}}.
\end{align}

We can determine the combination $\sum\limits_{i}a_{0}^{(i)}b_{-}^{(i)*}$ by multiplying both sides of (\ref{eft pi zero}) by $b_{-}^{(i)*}$ and summing over $i$.  This gives
\begin{align}
\label{first a0 b- eq}
\sum\limits_{i}a_{0}^{(i)}b_{-}^{(i)*} = \sum\limits_{i}c_{1}^{(i)}b^{(i)*}_{-} - \frac{3\mu H}{2\left(\mu^{2} + m^{2}\right)}.
\end{align}
We have already shown that $\sum\limits_{i}{\rm Re}\left(c_{1}^{(i)}b_{-}^{(i)*}\right) = 0$, which implies
\begin{align}
\label{IR constraints 1}
\sum\limits_{i}{\rm Re}\left(a_{0}^{(i)}b^{(i)*}_{-}\right) = -\frac{3\mu H}{2\left(\mu^{2} + m^{2}\right)}.
\end{align}
The remaining combinations of power series coefficients  needed to compute the higher order correlation functions of $\pi$ are fixed using the canonical commutation relations of $s$ and $\pi$. 
Consider the equal time relation $\left[s({\bf x},\tau), \pi({\bf y},\tau)\right] = 0$.  By inserting (\ref{eq:mode}) and (\ref{eq:mode2}) into this relation, we find
\begin{align}
\left[\pi({\bf x},\tau), s({\bf y},\tau)\right] = \int \frac{d^{3}{\bf k}}{(2\pi)^{3}}e^{i{\bf k}\cdot\left({\bf x} - \bf {y}\right)}\sum\limits_{i}\left[\pi^{(i)}_{k}(\eta)s_{k}^{(i)*}(\eta)  - {\rm c.c.} \right] = 0.
\end{align} 
The mode functions must then satisfy
\begin{align}
\label{comm:s phi}
\sum\limits_{i}{\rm Im}\left[\pi^{(i)}_{k}(\eta)s_{k}^{(i)*}(\eta)\right] = 0
\end{align}
for all $\eta$.  Plugging the leading IR behavior of the mode functions (\ref{small mode function behavior}) into (\ref{comm:s phi}) and demanding it holds at orders $(-\eta)^{\alpha_-}$, $(-\eta)^{\alpha_+}$, $(-\eta)^2$, and $(-\eta)^3$ respectively yields the following constraints
\begin{align}
\label{IRconstraints: phi s}
 \sum\limits_{i}{\rm Im}\left[a^{(i)}_{0}b^{(i)*}_{-}\right]
&= \sum\limits_{i}{\rm Im}\left[a^{(i)}_{0}b^{(i)*}_{+}\right] 
=\sum\limits_{i}{\rm Im}\left[a^{(i)}_{0}b^{(i)*}_{0,2}\right]\cr 
&= \sum\limits_{i}{\rm Im}\left[a^{(i)}_{0}b^{(i)*}_{3} + a^{(i)}_{+}b^{(i)*}_{-} + a^{(i)}_{-}b^{(i)*}_{+}\right] = 0. 
\end{align}
Given the fact that the recursion relations ({\ref{coefficient equations}}) and eq. (\ref{btoas}) are real, eqs. (\ref{IRconstraints: phi s}) and (\ref{btoas}) further imply that:
\begin{align}
\label{more zero comms}
\sum_i{\rm Im}\left[a_0^{(i)}b_{-,2}^{(i)*}\right]=\sum_i {\rm Im}\left[b_-^{(i)}b_{0,2}^{(i)*}\right]=0
\end{align}
Moreover, the recursion relations ({\ref{coefficient equations}) being real along with the fact that $\sum_i{\rm Im}\left[|b_-^{(i)}|^2\right]=0$ imply that
\begin{align}
\label{bb zero comm}
\sum_i {\rm Im}\left[b_-^{(i)}b_{-,2}^{(i)*}\right]=0
\end{align}
Furthermore, using the commutation relation $\left[\pi({\bf x},\tau), \Pi_{\pi}({\bf y},\tau)\right] = i\delta^{3}({\bf x} - {\bf y})$ gives:
\begin{align}
\label{phi Pi phi}
&\sum\limits_{i}{\rm Im}\left[3a^{(i)}_{0}a^{(i)*}_{3} + \alpha_{+}a^{(i)}_{-}a^{(i)*}_{+} + \alpha_{-}a^{(i)}_{+}a^{(i)*}_{-}\right] = -\frac{1}{2}\cr 
&\sum_i {\rm Im}\left[a_-^{(i)}a_3^{(i)*}\right]=0
\end{align}
Again using the fact that relations (\ref{btoas}) are real, we can convert the second equation in (\ref{phi Pi phi}) to:
\begin{align}
\label{bb3 zero comm}
\sum_i{\rm Im}\left[b_-^{(i)}b_3^{(i)*}\right]=0
\end{align}
Using (\ref{btoas}), we can combine the final equation of  (\ref{IRconstraints: phi s}) with the first equation of (\ref{phi Pi phi}) to find
\begin{align}
\label{IR constraints 2}
\sum\limits_{i}{\rm Im}\left[a^{(i)}_{0}b^{(i)*}_{3}\right] &= \frac{\mu H}{2\left(\mu^{2} + m^{2}\right)}\cr 
\sum\limits_{i}{\rm Im}\left[b_{-}^{(i)}b^{(i)*}_{+}\right] &= \frac{-1}{2\left(\alpha_{+} - \alpha_{-}\right)} \simeq -\frac{1}{6}.
\end{align}
The equalities in eq. (\ref{IR constraints 2}) hold for all $m$ and $\mu$ such that $m^{2} + \mu^{2} \leq 9H^{2}/4$, i.e. for $\alpha_{-}$ and $\alpha_{+}$ real. 

Equations (\ref{b minus squared}), (\ref{leading C2}), (\ref{IR constraints 1}), (\ref{more zero comms}), (\ref{bb zero comm}), (\ref{bb3 zero comm}), and (\ref{IR constraints 2}) comprise the full set of relations among power series coefficients we need to compute the leading $m$ and $\mu$ dependence of the correlation functions of $\pi$.  We will also need the fact that $n>0$ coefficients $a_{r,n}^{(i)}$ and $b_{r,n}^{(i)}$ are not enhanced by powers of $\alpha_-^{-1}$ compared to $a_r^{(i)}$ and $b_{r}^{(i)}$ coefficients for small $\alpha_-$, a fact which is simple to see from the recursion relations (\ref{coefficient equations}).

\section{Primordial Non-Gaussianities}
\label{Primordial Non-Gaussianities label}
In this section we compute the leading $m$ and $\mu$ behavior of the connected three- and four-point functions of the curvature perturbation $\zeta$ for arbitrary external wave-vectors.  We also compute the connected five-and six-point functions in certain kinematic limits.  We will use these results to calculate the two- and three-point functions of biased objects.  

We perform the computation of these correlation functions using the in-in formalism \cite{Weinberg:2005vy}.  We will mostly use the commutator form of the in-in correlator of an operator $\mathcal{O}(0)$: 
\begin{align}
\label{in in}
\langle \mathcal{O}(0) \rangle = \sum_{N=0}^{\infty}i^{N}\int_{-\infty}^{0}d\tau_{N}\int_{-\infty}^{\tau_{N}}d\tau_{N-1}\dots \int_{-\infty}^{\tau_{2}}d\tau_{1}\langle \lbrack H_{int}^{I}(\tau_{1}), \lbrack H_{int}^{I}(\tau_{2}),\dots\lbrack H_{int}^{I}(\tau_{N}),\mathcal{O}^{I}(0)\rbrack \dots \rbrack \rangle_{I}
\end{align}
where $I$ denotes a state or operator evolving in the interaction picture and $H_{int}$ denotes the interaction Hamiltonian\footnote{We restrict our attention to renormalizable terms in the potential for $s$.}
\begin{align}
\label{Hint}
H_{int}(\tau) = \frac{1}{(H \tau)^4}\int d^{3}{\bf x} \left[\frac{1}{\Lambda}s(x) g^{\mu\nu}\partial_{\mu}\pi(x) \partial_{\nu} \pi(x) + \frac{V'''}{3!}s(x)^{3} + \frac{V^{(4)}}{4!}s(x)^{4}\right].
\end{align}
 For simplicity, we assume $V^{(4)}$ is much smaller than $\left(V'''/H\right)^{2}$ and can be neglected.  We have also explored the importance of the $s\partial \pi \partial \pi$ interaction in comparison with the $s^3$ interaction for the primordial curvature bispectrum. For the range of parameters that we are using in this paper, we find numerically that the ratio of these contributions is $O(10^{-3})/f_{NL}$.  We suspect that this interaction is subdominant for the other primordial correlation functions as well, and neglect this interaction henceforth.  All relevant interactions are then mediated by the $V'''$ term.  We assume $|V'''|/H<1$ so that perturbation theory is valid.

\subsection{Three-Point Function}\label{three point function section}
The three-point function of $\zeta$ can be written
\begin{align}
\left<\zeta_{{\bf k}_{1}}\zeta_{{\bf k}_{2}}\zeta_{{\bf k}_{3}}\right> \equiv B_{\zeta}({\bf k}_{1},{\bf k}_{2},{\bf k}_{3})(2\pi)^{3}\delta^{3}\left({\bf k}_{1} + {\bf k}_{2} + {\bf k}_{3}\right).
\end{align}
The leading contribution to the bispectrum $B_{\zeta}({\bf k}_{1},{\bf k}_{2},{\bf k}_{3})$ is obtained by inserting a single factor of the $V'''$ interaction into (\ref{in in}).  This yields
\begin{align}
\label{3 point no normalization}
B_{\zeta}({\bf k}_{1},{\bf k}_{2},{\bf k}_{3}) &=-2V'''\left(\frac{H}{\dot{\phi}_{0}}\right)^{3}{\rm Im}\int\limits_{-\infty}^{0}\frac{d\tau}{(H\tau)^{4}}\prod_{l=1}^{3}\left[\pi^{(1)}_{k_{l}}(0)s_{k_{l}}^{(1)*}(k_{l} \tau) + \pi^{(2)}_{k_{l}}(0)s_{k_{l}}^{(2)*}(k_{l} \tau)\right].
\end{align}
Equation (\ref{3 point no normalization}), written in terms of the rescaled mode functions (\ref{rescaled mode functions}), becomes 
\begin{align}
\label{3 point}
B_{\zeta}({\bf k}_{1},{\bf k}_{2},{\bf k}_{3}) = -2\left(\frac{H^{2}}{\dot{\phi}_{0}}\right)^{3}\left(\frac{V'''}{H}\right)\left(\prod_{i}^{3}\frac{1}{k_{i}^{3}}\right){\rm Im}\int\limits_{-\infty}^{0}\frac{d\tau}{\tau^{4}}\prod_{l=1}^{3}\sum\limits_{i}\pi^{(i)}(0)s^{(i)*}(k_{l} \tau).
\end{align}
Let us now focus on the evaluation of the integral in (\ref{3 point}), which can be written:
\begin{align}
\label{3 point master integral}
k_{ UV}^{3}{\rm Im}\int\limits_{-\infty}^{0}\frac{d\eta}{\eta^{4}}\prod_{l =1}^{3}\sum\limits_{i}\pi^{(i)}(0)s^{(i)*}\left(\frac{k_{l}}{k_{UV}}\eta\right)
\end{align}
where we define $k_{UV} = {\rm max}(k_{l})$ and $\eta=k_{UV}\tau$.  In the small $\mu$ and $m$ regime, (\ref{3 point master integral}) receives most of its support from the IR region of the integral (when the arguments of the mode functions are less than 1 in magnitude) due to the superhorizon growth mentioned in the discussion following (\ref{small tau pi in terms of s}).  The contribution from the UV region is subdominant.  Our choice of $k_{UV}$ implies the leading $m$ and $\mu$ contribution to the integral comes from the region $-1 \le \eta \le 0$, and (\ref{3 point master integral}) becomes:
\begin{align}
\label{IR 3 point}
&k_{UV}^{3}{\rm Im} \int\limits_{-1}^{0}\frac{d\eta}{\eta^{4}}\prod_{l=1}^{3}\sum\limits_{i}\left[\left(a^{(i)}_{0}b^{(i)*}_{-}\right)\left(-\frac{k_{l}}{k_{UV}}\eta\right)^{\alpha_{-}} + \left(a^{(i)}_{0}b^{(i)*}_{0,2}\right)\left(-\frac{k_{l}}{k_{UV}}\eta\right)^{2}\right.\cr
&\left. + \left(a^{(i)}_{0}b^{(i)*}_{-,2}\right)\left(-\frac{k_{l}}{k_{UV}}\eta\right)^{\alpha_{-}+2} +\left(a^{(i)}_{0}b^{(i)*}_{+}\right)\left(-\frac{k_{l}}{k_{UV}}\eta\right)^{\alpha_{+}} + \left(a^{(i)}_{0}b^{(i)*}_{3}\right)\left(-\frac{k_{l}}{k_{UV}}\eta\right)^{3} + O(\eta^4) \right].\cr
\end{align}
Note the integral is potentially IR divergent because of the factor of $1/\eta^{4}$.  However, eqs. (\ref{IRconstraints: phi s}) and (\ref{more zero comms}) imply the coefficients multiplying the IR divergent terms are zero, and that the leading $\mu$ and $m$ behavior of (\ref{IR 3 point}) is
\begin{align}
\label{C3 equil IR approx}
\left(\sum\limits_{i} {\rm Re}\left[a^{(i)}_{0}b_{-}^{(i)*}\right]\right)^{2} & \left(\sum\limits_{i}{\rm Im}\left[a^{(i)}_{0}b^{(i)*}_{3}\right]\right)\left[k^{3}_{1}\left(\frac{k_{2}k_{3}}{k_{UV}^{2}}\right)^{\alpha_{-}} + {\rm cyc.\ perm}\right]\int\limits_{-1}^{0}d\eta (-\eta)^{-1 + 2\alpha_{-}}\cr
&= \frac{27}{16}\frac{\mu^{3}H^5}{\left(\mu^{2} + m^{2}\right)^{4}}\left[k^{3}_{1}\left(\frac{k_{2}k_{3}}{k_{UV}^{2}}\right)^{\alpha_{-}} + {\rm cyc.\ perm} \right].
\end{align}
As long as $\alpha_{-}$ is small, the answer does not depend on the precise choice of $k_{UV}$, we only have to choose it to be of the same order as the hardest wave-vector entering the vertex.\footnote{The ratios of external wave-vectors to $k_{UV}$ raised to the power $\alpha_{-}$ in equation (\ref{C3 equil IR approx}) can be interpreted as the re-summation of leading logs in the $\alpha_{-}$ expansion.}   Equivalently, the answer is insensitive to the precise choice of the lower bound of the $\eta$ integral.  Plugging (\ref{C3 equil IR approx}) into (\ref{3 point}), we find that the leading $m$ and $\mu$ behavior of the $O(V''')$ contribution to the bispectrum is
\begin{align}
\label{3 point leading behavior}
B_{\zeta}({\bf k}_{1},{\bf k}_{2},{\bf k}_{3}) &= -\left(\frac{H^{2}}{\dot{\phi}_{0}}\right)^{3}\left(\frac{V'''}{H}\right)\frac{1}{k^{3}_{1}k^{3}_{2}k^{3}_{3}} \frac{\left(3\mu/2\right)^{3}H^5}{\left(\mu^{2} + m^{2}\right)^{4}}\cr
&\ \ \ \ \ \ \ \times\left[k^{3}_{1}\left(\frac{k_{2}k_{3}}{k_{UV}^{2}}\right)^{\alpha_{-}} + k^{3}_{2}\left(\frac{k_{1}k_{3}}{k_{UV}^{2}}\right)^{\alpha_{-}} + k^{3}_{3}\left(\frac{k_{1}k_{2}}{k_{UV}^{2}}\right)^{\alpha_{-}}\right]
\end{align}
where $k_{UV} = {\rm max}(k_{i})$.  Equation (\ref{3 point leading behavior}) was computed numerically in \cite{Chen:2009zp} and is valid for any external wave-vector configuration.  Note that when the wave-vectors $k_1$, $k_2$ and $k_3$ are  the same order of magnitude, the terms raised to the power $\alpha_{-}$ can be set to unity. Then the bispectrum has the same form as local non-Gaussianity, {\it i.e.}, $B_{\zeta}({\bf k}_{1},{\bf k}_{2},{\bf k}_{3}) \propto \left[P_{\zeta}(k_1)P_{\zeta}(k_2)+P_{\zeta}(k_1)P_{\zeta}(k_3)+P_{\zeta}(k_2)P_{\zeta}(k_3)\right]$.

\begin{figure}
\includegraphics[width=2in]{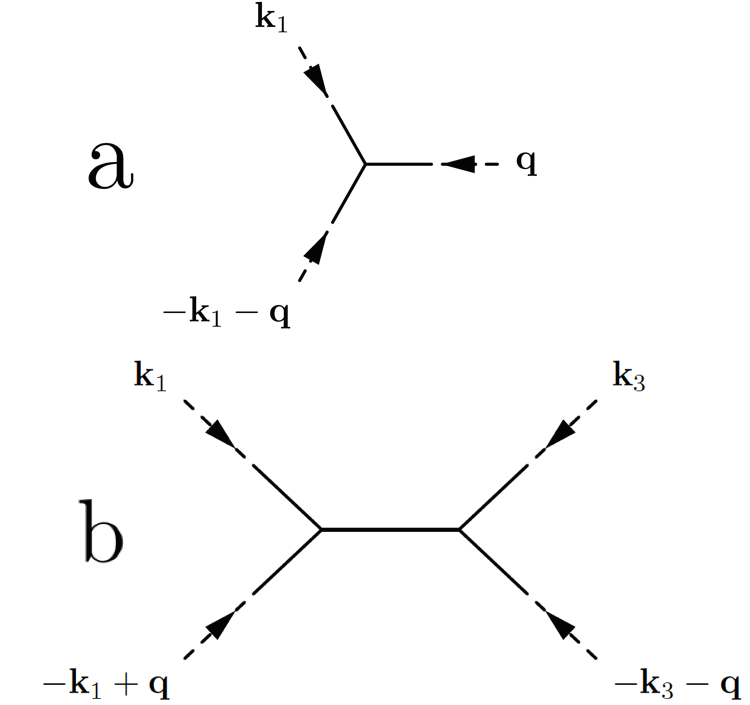}
\caption{Diagrams that contribute to three- and four-point correlations of $\zeta$ in the squeezed and collapsed limits respectively.  These diagrams contribute to the galactic halo power spectrum. Dashed lines represent $\pi$, while solid lines represent $s$.}
\label{fig:feynman 2 pt}
\end{figure}

We now study (\ref{3 point leading behavior}) in a couple interesting kinematic limits.  First, consider (\ref{3 point leading behavior}) in the equilateral limit $k_{i} \equiv k$
\begin{align}
\label{3 point equil B}
B^{\rm equil}_{\zeta}(k) = -\left(\frac{H^{2}}{\dot{\phi}_{0}}\right)^3\left(\frac{V'''}{H}\right)\frac{1}{k^{6}}\frac{3\left(3\mu/2\right)^{3}H^5}{\left(\mu^{2} + m^{2}\right)^{4}}.
\end{align}
We can use (\ref{3 point equil B}) to relate $V'''$ to the model's prediction for $f_{NL}$.  We estimate $f_{NL}$ using
\begin{align}
\label{fequil}
f_{NL} = \frac{5}{18} \times \frac{B^{{\rm equil}}_{\zeta}(k)}{\mathcal{P}_{\zeta}(k)^{2}}.
\end{align}   
Substituting (\ref{power spectrum}), (\ref{phi not dot}) and (\ref{3 point equil B}) into (\ref{fequil}) gives
\begin{align}
\label{fnl bound on V'''}
\frac{V'''}{H} = -\frac{6}{5}f_{NL}\sqrt{2\pi^{2}\Delta^{2}_{\zeta}}C_{2}(\mu,m)^{\frac{3}{2}}\frac{(\mu^2+m^2)^4}{(3\mu/2)^3H^5}. 
\end{align} 
The current Planck $95\%$ C.L. constraint for local non-Gaussianity is $f_{NL}=2.7\pm 11.6$.  For $f_{NL}=10$ and $\mu/H=m/H=0.3$, we find that $|V'''|/H\simeq 10^{-3}$.

The two kinematic configurations we will be most interested in when we compute galactic halo correlators are when all three external wave-vectors are soft (Fig. \ref{fig:feynman 3 pt}$c$), and when one leg is soft while the other two are hard - the so-called squeezed limit (Fig. \ref{fig:feynman 2 pt}$a$).  In what follows, we will denote hard wave-vectors by $k$ and soft wave-vectors by $q$.  First we consider the squeezed limit.  We choose ${\bf k}_{2} = -{\bf k}_{1} - {\bf q}$ and $k_{1} = k_{2} \equiv k >> k_{3} \equiv q$.  The full $O(V''')$ contribution (\ref{3 point}) to the bispectrum in this limit can be written
\begin{align}
\label{B squeezed}
B^{\rm sq}_{\zeta}(k,q) =-\left(\frac{H^{2}}{\dot{\phi}_{0}}\right)^3\left(\frac{V'''}{H}\right)\frac{2\left(3\mu/2\right)^{3}H^5}{\left(\mu^{2} + m^{2}\right)^{4}}\frac{1}{k^{3+\alpha_{-}}q^{3-\alpha_{-}}}.
\end{align}
The wave-vector dependence of equation (\ref{B squeezed}) was first determined in \cite{Chen:2009zp,Sefusatti:2012ye}.  Finally, the bispectrum in the limit where all three external wave-vectors are soft can be obtained simply by making the replacement $k_{i} \rightarrow q_{i}$ in (\ref{3 point leading behavior}).

\subsection{Four-Point Function}
\label{Four Point Function label}
The four-point function of $\zeta$ can be written
\begin{align}
\label{4 point def}
\left<\zeta_{{\bf k}_{1}}\zeta_{{\bf k}_{2}}\zeta_{{\bf k}_{3}}\zeta_{{\bf k}_{4}}\right> \equiv N^{(4)}_{\zeta}({\bf k}_{1},{\bf k}_{2},{\bf k}_{3},{\bf k}_{4})(2\pi)^{3}\delta^{3}\left(\sum_{i}^{4}{\bf k}_{i}\right).
\end{align}
We can derive the leading contribution to $N_{\zeta}^{(4)}$ by inserting two factors of the $V'''$ interaction into (\ref{in in}).  It is convenient to define
\begin{align}
A(x)\equiv \sum_{i}\pi^{(i)}(0)s^{(i)*}(x)\ \ \ \ \ \ \ B(x)\equiv \sum_{i}b_-^{(i)}s^{(i)*}(x).
\end{align}
By expanding the commutators and performing all possible contractions, we find:
\begin{align}
\label{4 point full}
&N^{(4)}_{\zeta}({\bf k}_{1},{\bf k}_{2},{\bf k}_{3},{\bf k}_{4}) = 4 \left(\frac{H^{2}}{\dot{\phi}_{0}}\right)^{4}\left(\frac{V'''}{H}\right)^{2}\left(\prod_{i}^{4}\frac{1}{k_{i}^{3}}\right)\frac{1}{k_{12}^{3}}\int\limits_{-\infty}^{0}\frac{d\tau}{\tau^{4}}\int\limits_{-\infty}^{\tau}\frac{d\tau'}{\tau'^{4}}\cr
&\ \ \ \ \ \ \ \ \times{\rm Im}\left[A(k_1 \tau)A(k_2 \tau) \right]{\rm Im}\left[A(k_3 \tau')A(k_4 \tau')\sum\limits_{i}s^{(i)}(k_{12}\tau)s^{(i)*}(k_{12}\tau')\right]\cr 
&\ \ \ \ \ \ \ \ \ \ \ \ + (k_{1} \leftrightarrow k_{3}, k_2 \leftrightarrow k_4) + {\rm cyc. \ perms}({\bf k}_{2},{\bf k}_{3},{\bf k}_{4})
\end{align}
where $k_{12} =\left |{\bf k}_{1} + {\bf k}_{2}\right|$. 

Unlike the calculation of the three-point function, the four-point one involves nested time integrals.  Again, the four-point integral is dominated by the IR for $\alpha_{-} << 1$ and the integrand reduces to polynomials in $\tau$ and $\tau'$.  Like before, we make the change of variable $\eta\equiv k_{UV_{12}}\tau$ and $\eta'\equiv k_{UV_{34}}\tau'$, where $k_{UV_{ij}}\equiv \text{max}\{k_i,k_j,|{\bf k}_i+{\bf k}_j|\}$ and cut off the integrals at $\eta_{UV}=-1$ and $\eta_{UV}'=-1$ (recall that the result is not sensitive to this cutoff value as long as $\alpha_{-}$ is small).  The relationships among the power series coefficients deduced in section \ref{commutation relations section and other stuff} imply the integral converges in the IR.  

Without loss of generality, assume that $k_1$ is the largest external wave-vector (this implies that $k_{UV_{12}}\ge k_{UV_{34}}$).  Using the identities relating the power series coefficients derived in section \ref{commutation relations section and other stuff}, the time integral in (\ref{4 point full}) becomes:
\begin{align}
\label{first term 4 pt final}
&\int\limits_{-\infty}^{0}\frac{d\tau}{\tau^{4}}\int\limits_{-\infty}^{\tau}\frac{d\tau'}{\tau'^{4}}{\rm Im}\left[A(k_1 \tau)A(k_2 \tau)\right]{\rm Im}\left[A(k_3\tau')A(k_4 \tau')\sum\limits_{i}s^{(i)}(k_{12}\tau)s^{(i)*}(k_{12}\tau')\right] + \left(k_{1} \leftrightarrow k_{3}, k_2 \leftrightarrow k_4\right)\cr
&= \frac{9}{32}\frac{\mu^4H^4}{\left(\mu^{2} + m^{2}\right)^4}\left[\left(\frac{k_{I}^2}{k_{UV_{12}}^2k_{UV_{34}}^2}\right)^{\alpha_{-}}\left[k_{1}^{3}k_{2}^{\alpha_{-}} + k_{2}^{3}k_{1}^{\alpha_{-}}\right]\left[k_{3}^{3}k_{4}^{\alpha_{-}} + k_{4}^{3}k_{3}^{\alpha_{-}}\right]\right.\cr
&\ \ \ \ \times \left[\int\limits_{-1}^{0}d\eta(-\eta)^{-1 + 2\alpha_{-}}\int\limits_{-1}^{\frac{k_{UV_{34}}}{k_{UV_{12}}}\eta}d\eta' (-\eta')^{-1 + 2\alpha_{-}} + \int\limits_{-\frac{k_{UV_{34}}}{k_{UV_{12}}}}^{0}d\eta(-\eta)^{-1 + 2\alpha_{-}}\int\limits_{-1}^{\frac{k_{UV_{12}}}{k_{UV_{34}}}\eta}d\eta' (-\eta')^{-1 + 2\alpha_{-}}\right]\cr 
& +\left(\frac{k_I^3}{{k_{UV_{12}}}^{2\alpha_-}{k_{UV_{34}}}^{\alpha_-}}\right)\left[k_{1}^{3}k_{2}^{\alpha_{-}} + k_{2}^{3}k_{1}^{\alpha_{-}}\right]\left[k_{3}^{\alpha_-}k_{4}^{\alpha_{-}} \right]\int\limits_{-1}^{0}d\eta(-\eta)^{-1 + 2\alpha_{-}}\int\limits_{-1}^{\frac{k_{UV_{34}}}{k_{UV_{12}}}\eta}d\eta' (-\eta')^{-1 + \alpha_{-}} \cr 
&\left. +\left(\frac{k_I^3}{{k_{UV_{12}}}^{\alpha_-}{k_{UV_{34}}}^{2\alpha_-}}\right)\left[k_{1}^{\alpha_-}k_{2}^{\alpha_{-}}\right]\left[k_{3}^{3}k_{4}^{\alpha_-}+k_{3}^{\alpha_-}k_{4}^{3}\right] \int\limits_{-\frac{k_{UV_{34}}}{k_{UV_{12}}}}^{0}d\eta(-\eta)^{-1 + 2\alpha_{-}}\int\limits_{-1}^{\frac{k_{UV_{12}}}{k_{UV_{34}}}\eta}d\eta' (-\eta')^{-1 + \alpha_{-}}\right].
\end{align}
Notice not all of the lower bounds of the $\eta$ integrals equal -1, some are cutoff by $-\frac{k_{UV_{34}}}{k_{UV_{12}}}$.  This is to ensure that the upper bound of the $\eta'$ integral is greater than -1.  Evaluating the time integrals, we find the four-point function for general external wave-vectors is
\begin{align}
\label{small mu m 4 pt}
&N_\zeta^{(4)}({\bf k}_1,{\bf k}_2,{\bf k}_3,{\bf k}_4)=\left(\frac{H^2}{\dot \phi_0}\right)^4\left(\frac{V'''}{H}\right)^2\left(\prod_{i=1}^4\frac{1}{k_i^3}\right)\frac{1}{k_{12}^3}\frac{(3\mu/2)^4H^8}{2(\mu^2+m^2)^6}\cr 
&\ \ \ \times\left[\left(k_1^3 k_2^{\alpha_-}+k_1^{\alpha_-}k_2^3\right)\left(k_3^3 k_4^{\alpha_-}+k_3^{\alpha_-}k_4^3\right)\left(\frac{k_{12}}{k_{UV_{12}}k_{UV_{34}}}\right)^{2\alpha_-}\right.\cr 
&\ \ \ \ \ \ \ \ \ \ \ +2\left(1-\frac{2}{3}\left(\frac{k_{UV_{34}}}{k_{UV_{12}}}\right)^{\alpha_-}\right)(k_1^3 k_2^{\alpha_-}+k_1^{\alpha_-}k_2^3)(k_3 k_4)^{\alpha_-}\frac{k_{12}^3}{k_{UV_{12}}^{2\alpha_-}k_{UV_{34}}^{\alpha_-}}\cr
&\ \ \ \ \ \ \ \ \ \ \ \ \ \ \ \ \ \left.+\frac{2}{3}\left(k_1 k_2\right)^{\alpha_-}\left(k_3^3k_4^{\alpha_-}+k_3^{\alpha_-}k_4^3\right)\frac{k_{12}^3}{k_{UV_{12}}^{3\alpha_-}}\right]+\text{cyc. perm}({\bf k}_2,{\bf k}_3,{\bf k}_4) 
\end{align}

\begin{figure}
\includegraphics[width=5in]{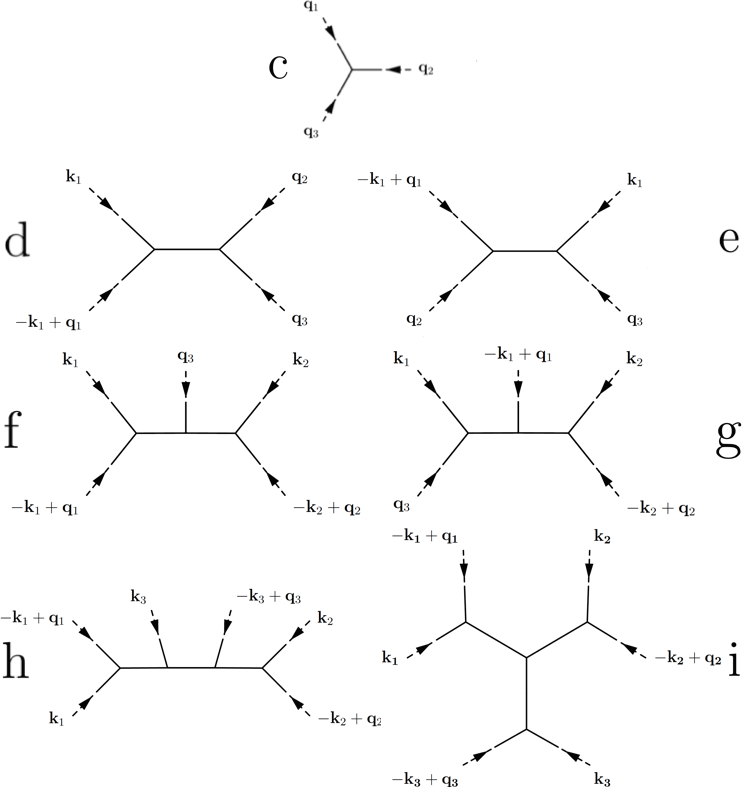}
\caption{Diagrams that contribute to the three-, four-, five-, and six-point correlations of $\zeta$ in the kinematic regimes that contribute to the enhanced part of the galactic halo bispectrum.  Dashed lines represent $\pi$, while solid lines represent $s$.}
\label{fig:feynman 3 pt}
\end{figure}

We now focus on kinematic limits of (\ref{small mu m 4 pt}) that are most important in the calculation of the two- and three-point functions of galactic dark matter halos.  The enhancements discovered in~\cite{AGW} and~\cite{Dalal:2007cu} respectively occur when the magnitude of a sum of wave-vectors in the correlation function of $\zeta$ is small or when the magnitude of an external wave-vector is small. For the four-point correlation, the first of these is referred to as the collapsed limit. Suppose that $q$ denotes small wave-vectors, and $k$ denotes large wave-vectors. In these computations (as well as in later computations of the five- and six-point functions of $\zeta$), we assume that $(k_i/k_j)^{\alpha_-}\simeq 1$ and $(q_i/q_j)^{\alpha_-}\simeq 1$.  This approximation is justified in our application to galactic halos since we will want to consider $k$'s roughly on the order of the inverse of the galactic halo radius, and since the $q$'s will be taken to be within an order of magnitude from each other (i.e. between about $(50~{\rm Mpc}/h)^{-1}$ and $(1000~ {\rm Mpc}/h)^{-1}$).  However, we do not take $(q/k)^{\alpha_-}$ to be approximately $1$ since $q$ and $k$ may differ by several orders of magnitude.  We first specialize to the collapsed limit of (\ref{small mu m 4 pt}) which occurs when two pairs of legs have nearly equal and opposite wave-vectors.  Let ${\bf k}_{2} = -{\bf k}_{1}+{\bf q}$ and ${\bf k}_{4} = -{\bf k}_{3}-{\bf q}$ where $q<<k_1,\ k_3$.  Then the most important permutation of (\ref{small mu m 4 pt}) in this collapsed limit is when ${\bf k}_{1}$ and ${\bf k}_{2}$ are attached to one vertex, and ${\bf k}_{3}$ and ${\bf k}_{4}$ are attached to the other.  The wave-vector of the internal line becomes very small (Fig. {\ref{fig:feynman 2 pt}$b$) and eq.~(\ref{small mu m 4 pt}) becomes
\begin{align}
\label{analytic 4 point comp}
N^{(4),\text{ coll}}_{\zeta}({\bf k}_{1},-{\bf k}_{1}+{\bf q},{\bf k}_{3},-{\bf k}_{3}-{\bf q}) = \left(\frac{H^{2}}{\dot{\phi}_{0}}\right)^{4}\left(\frac{V'''}{H}\right)^{2}\frac{1}{q^{3-2\alpha_{-}}}\frac{1}{(k_{1}k_{3})^{3+\alpha_{-}}}\frac{2(3\mu/2)^4H^8}{(\mu^2+m^2)^6}.
\end{align}
The four-point in the collapsed limit was previously computed in~\cite{Assassi:2012zq}.

The other interesting kinematic limit of (\ref{small mu m 4 pt}) is when one pair of legs have nearly equal and opposite wave-vectors and the wave-vectors of the other two legs are soft.  We find for the sum of Figs. \ref{fig:feynman 3 pt}$d$ and \ref{fig:feynman 3 pt}$e$:
\begin{align}
&N^{(4)}_{\zeta}({\bf k}_1,-{\bf k}_1+{\bf q}_{1},{\bf q}_{2},{\bf q}_{3}) =\left(\frac{H^{2}}{\dot{\phi}_{0}}\right)^{4}\left(\frac{V'''}{H}\right)^{2}\frac{(3\mu/2)^{4}H^8}{\left(\mu^{2} + m^{2}\right)^{6}}\frac{1}{k_1^3}\left(\frac{q}{k_{1}} \right)^{\alpha_{-}}\cr 
&\ \ \ \ \ \ \ \ \ \ \ \ \ \ \times\left[\frac{1}{q_1^3 q_2^3}+\frac{1}{q_1^3 q_3^3}+2\left(1+\frac{1}{2}\left(\frac{q}{k_1}\right)^{\alpha_-}\right)\frac{1}{q_2^3 q_3^3}\right].
\end{align}

\subsection{Five- and Six-Point Functions}

Given the techniques we have developed so far, it is possible to compute the five- and six-point functions of $\zeta$ for general external wave-vectors.  However, our primary purpose in studying these objects is to compute their most important contributions to the three-point function of galactic dark matter halos in the limit of large halo separation.  We then only focus on the kinematic limits of the five- and six-points giving rise to the largest long wavelength enhanced terms.  Even in these limits, the calculation is too long to present here.  In this section we just quote results and relegate an outline of the derivation to Appendix \ref{appendix 5 and 6}.

The strongest long wavelength enhanced behavior of the five-point function is achieved when one leg is soft and the other four come in pairs of nearly equal and opposite wave-vectors.  Panels $f$ and $g$ of Fig. \ref{fig:feynman 3 pt} illustrate this kinematic setup.  The contribution of these graphs to the five-point function is: 
\begin{align}
N_\zeta^{(5)}&({\bf k}_1,{\bf q}_1-{\bf k}_1,{\bf k}_2,{\bf q}_2-{\bf k}_2,{\bf q}_3)=-\left(\frac{H^2}{\dot \phi_0}\right)^5\left(\frac{V'''}{H}\right)^3\frac{(3\mu/2)^5H^{11}}{(\mu^2+m^2)^8}\frac{1}{k_1^3 k_2^3}\left(\frac{q^2}{k_1 k_2}\right)^{\alpha_-}\cr
&\ \ \ \ \ \ \ \times\left[\frac{1}{q_1^3 q_2^3}
+\left(2-\frac{1}{6}\left(\frac{q}{k_2}\right)^{\alpha_-}\right)\frac{1}{q_2^3 q_3^3}+\left(2-\frac{1}{6}\left(\frac{q}{k_1}\right)^{\alpha_-}\right)\frac{1}{q_1^3 q_3^3}\right]
\end{align}
where we have defined $q=\max\{q_i\}$.  

The most important long wavelength contributions to the six-point function occur when all six legs come in pairs of nearly equal and opposite wave-vectors.  The most important diagrams are displayed in panels $h$ and $i$ of Fig.  \ref{fig:feynman 3 pt} and the sum of their contributions is
\begin{align}
&N_\zeta^{(6)}({\bf k}_1,{\bf q}_1-{\bf k}_1,{\bf k}_2,{\bf q}_2-{\bf k}_2,{\bf k}_3,{\bf q}_3-{\bf k}_3)=\left(\frac{H^2}{\dot \phi_0}\right)^6\left(\frac{V'''}{H}\right)^4\frac{1}{k_1^3 k_2^3 k_3^3}\frac{2(3\mu/2)^6H^{14}}{(\mu^2+m^2)^{10}}\cr 
&\ \ \ \ \ \ \ \ \ \ \ \ \ \ \times \left(1+\frac{1}{2}\left(\frac{q^3}{k_1 k_2 k_3}\right)^{\alpha_-/3}\right)\left(\frac{q_1 q_2 q_3}{k_1 k_2 k_3}\right)^{\alpha_-}\left[\frac{1}{q_2^3q_3^3}+\frac{1}{q_1^3q_3^3}+\frac{1}{q_1^3q_2^3}\right]
\end{align}

\section{Correlation Functions of Biased Objects}

In this section we review the computation of the galactic halo power spectrum, and compute the bispectrum in the limit of large halo separation.  At large enough separation, the primordial non-Gaussian contributions to the power spectrum and bispectrum are larger than the Gaussian ones.  This leads to interesting observable long wavelength effects. The long wavelength scaling of the power spectrum was already discussed in ~\cite{Baumann:2012bc}.  Here we compute the long wavelength enhanced contributions and present results for the bispectrum as well.

We start by assuming halos form instantaneously, at the same time $t_{\rm coll}$, and at points where the matter overdensity $\delta({\bf x})$ averaged over a spherical region with comoving radius $R$ exceeds a threshold $\delta_c$.  We choose the smoothing radius $R$ to be of order the characteristic length scale of the region of space that collapses to form a halo.\footnote{We set $R = 2.8~ {\rm Mpc}.$}  The smoothed matter overdensity is related to the matter overdensity by
\begin{align}
\delta_{R}({\bf x},a) =\int d^{3}{\bf y}\ W_{R}(|{\bf x} - {\bf y}|)\delta({\bf y},a). 
\end{align}
Here $W_{R}(|{\bf x} - {\bf y}|)=\Theta_{H}(R-|{\bf x} - {\bf y}|)$ is the top hat window function.\footnote{$\Theta_{H}$ is the Heaviside step function.}  The Fourier transform of the window function is:
\begin{align}
\label{window function}
W_{R}(k) = \frac{3\left({\rm sin}kR - kR{\rm cos}kR \right)}{(kR)^{3}}.
\end{align}
Assuming $\delta({\bf x},a)$ undergoes linear growth before the collapse time, we can express the density perturbations at the time of collapse in terms of  the linearly evolved density perturbations today, $\delta_{R}({\bf x}$, $a_{\rm coll}) =\delta_{R}({\bf x})D(a_{\rm coll})$ where today $a=1$ and the growth factor $D(1)=1$.  

We will ignore the evolution of halos after collapse, and so the number density of halos today, up to an irrelevant dimensionful normalization constant, is given by:
\begin{align}
\label{halo formation model}
n_{h}({\bf x})  = \Theta_{H}(\delta_{R}({\bf x}, a_{\rm coll}) - \delta_{c}(a_{\rm coll})) = \Theta_{H}(\delta_{R}({\bf x}) - \delta_{c}) 
\end{align}
where $\delta_{c} \equiv \delta_{c}(a_{\rm coll})/D(a_{\rm coll})$.  We use $\delta_c=4.215$, which assumes that $\delta_c(a_{\rm coll})=1.686$ with $z_{\rm coll}=1.5$~\cite{{Press:1973iz}}.  The halo overdensity $\delta_{h}({\bf x})$ at a point ${\bf x}$ today is defined by
\begin{align}
\label{pertubation average density}
\delta_{h}({\bf x}) = \frac{n_{h}({\bf x}) - \left<n_{h}\right>}{\left<n_{h}\right>}.
\end{align}
where $\left<n_{h}\right>$ is the average halo density.

We are interested in the two- and three-point functions of $\delta_{h}({\bf x})$.  These can be computed using (\ref{halo formation model}) and the path integral techniques discussed in \cite{Politzer:1984nu}.  A more general approach that we adopt here is to write $\delta_h$ as\footnote{The ellipses denote higher order terms in the bias expansion. They are not needed to the order we work in $(qR)$ and $(V'''/H)$. However it is important to remember that they are defined with subtractions. For example, the next order term is $b_{3} (\delta_{R}^3({\bf x})-3\langle \delta_{R}^2 \rangle \delta_R({\bf x}))$.}$^{,}$\footnote{A completely general approach is possible; for a review, see \cite{Desjacques:2016bnm}.}
\begin{equation}
\label{bias expansion}
\delta_{h}({\bf x}) = b_{1} \delta_{R} ({\bf x}) + b_{2} (\delta_{R}^2({\bf x})-\langle \delta_{R}^2 \rangle )  + \dots
\end{equation}
 The constants $b_{1}$ and $b_{2}$ are bias coefficients.  They can be computed using a specific model of halo formation such as (\ref{halo formation model}) that expresses the halo overdensity in terms of  $\delta_R$ or determined from data. The two-point function of the halo overdensity is then:
\begin{eqnarray}
\label{2 point bias expan}
\left<\delta_{h}({\bf x})\delta_{h}({\bf y})\right> &=& b_{1}^{2}\left<\delta_{R}({\bf x})\delta_{R}({\bf y})\right>   \\ \nonumber
&&+b_{1}b_{2} \left< (\delta_{R}^{2}({\bf x})-\left<\delta_R^2\right>)\delta_{R}({\bf y})\right> + \left<\delta_{R}({\bf x})(\delta_{R}^{2}({\bf y})-\left<\delta_R^2\right>)\right> \\ \nonumber
&&+b_{2}^{2}\left<(\delta_{R}^{2}({\bf x})-\left<\delta_R^2\right>)(\delta_{R}^{2}({\bf y})-\left<\delta_R^2\right>)\right> + \dots ~.
\end{eqnarray}
Note
\begin{equation}
\left<\delta_{R}^{2}({\bf x})\delta_{R}^{2}({\bf y})\right>=\left<\delta_{R}^2\right>^{2} + \left<\delta_{R}({\bf x})\delta_{R}({\bf y})\right>^2+\left<\delta_{R}^{2}({\bf x})\delta_{R}^{2}({\bf y})\right>_c.
\end{equation}
We can neglect the second term because it is very small at large halo separations compared to the $b_1^2$ term in (\ref{2 point bias expan}).  All factors of $\left<\delta_{R}^2\right>$ cancel and we find 
\begin{equation}
\label{2 point bias expan1}
\left<\delta_{h}({\bf x})\delta_{h}({\bf y})\right> \simeq b_{1}^{2}\left<\delta_{R}({\bf x})\delta_{R}({\bf y})\right> +
b_{1}b_{2} (\left< \delta_{R}^{2}({\bf x})\delta_{R}({\bf y})\right> + \left<\delta_{R}({\bf x})\delta_{R}^{2}({\bf y})\right>) 
+b_{2}^{2}\left<\delta_{R}^{2}({\bf x})\delta_{R}^{2}({\bf y})\right>_c + \dots ~.
\end{equation}
The term proportional to $b_1^2$ comes from the Gaussian two-point function of $\zeta$ and the remaining terms arise from the connected three- and four-point functions of $\zeta$ that we computed earlier.

Similarly, we can express the three-point function of $\delta_{h}$ as:
\begin{eqnarray}
\label{biashalo3}
&&\left<\delta_{h}({\bf x})\delta_{h}({\bf y})\delta_{h}({\bf z})\right> =b_{1}^{3}\left<\delta_{R}({\bf x})\delta_{R}({\bf y})\delta_{R}({\bf z})\right>_c +b_{2}^{3}\left<\delta({\bf x})^{2}\delta({\bf y})^{2} \delta({\bf z})^{2}\right>_c \nonumber \\
&&\ \ \ \ \ \ +\left[2b_{1}^{2}b_{2}\left<\delta_{R}({\bf x})\delta_{R}({\bf y})\right>\left<\delta_{R}({\bf x})\delta_{R}({\bf z})\right>+b_{1}^{2}b_{2}\left<\delta_{R}^{2}({\bf x})\delta_{R}({\bf y})\delta_{R}({\bf z})\right>_c \right. \nonumber  \\ 
&&\left.\ \ \ \ \ \ \ \ \ \ \ \ \ \  + b_{1}b_{2}^{2}\left<\delta_{R}^{2}({\bf x})\delta_{R}^{2}({\bf y})\delta_{R}({\bf z})\right>_c + {\rm cyc. \ perm}({\bf x},{\bf y},{\bf z})\right] +\ldots~ 
\end{eqnarray}
 The first term proportional to $b_1^2 b_2$ is the three-point halo correlation when the underlying curvature perturbations are Gaussian, which was first calculated in~\cite{Politzer:1984nu}. The remaining terms arise from the non-Gaussian correlations of the primordial fluctuations.
In the next section we present a power counting argument showing that for widely separated points $|{\bf x}-{\bf y}|>> R$ and $|V'''|/H <1$, the higher order terms in the bias expansion are negligible in the threshold model. Only $b_1$ and $b_2$ are needed to compute the halo overdensity power spectrum and bispectrum evaluated at wave-vectors $q<<1/R$.

Using, for example, path integral methods, it is straightforward to derive expressions for $\left<n_h\right>$ and the bias coefficients $b_1$ and $b_2$ in the threshold model mentioned above. They can be expressed in terms of $\delta_c$ and 
\begin{equation}
\sigma_R^2=\langle \delta_R({\bf x})\delta_R({\bf x})\rangle
\end{equation}
as
\begin{equation}
\left< n_h \right>={1 \over 2}{\rm erfc}\left( \delta_c \over \sqrt{2} \sigma_R \right)
\end{equation}
and
\begin{equation}
\label{bias points above threshold}
b_1={e^{-\delta_c^2/(2\sigma_R^2)} \over \sqrt{2 \pi} \sigma_R \left<n_h\right>}~~~~b_2={e^{-\delta_c^2/(2\sigma_R^2)} \delta_c\over 2 \sqrt{2 \pi} \sigma_R^3\left<n_h\right>}.
\end{equation}
The Fourier transformed smoothed matter overdensity $\delta_{R}({\bf k})$ is related to the curvature perturbation through
\begin{align}
\label{delta to xi main text}
\delta_{R}({\bf k}) = \frac{2k^2}{5\Omega_m H_0^2}T(k)W_R(k)\zeta_{{\bf k}}
\end{align}
where $T(k)$ is the transfer function, $\Omega_{m}$ is the ratio of the matter density to the critical density today, and $H_{0}$ is the Hubble constant evaluated today \cite{Dodelson:2003ft}.  
When performing integrals against $T(k)$ we use the BBKS approximation to the transfer function \cite{Bardeen:1985tr}:
\begin{align}
\label{transfer function}
T\left(k=\left(\Omega_{m}h^{2}{\rm Mpc^{-1}}\right)u\right) = \frac{{\rm ln}\left[1 + 2.34 u\right]}{(2.34 u)}\left[1 + 3.89u + (16.2 u)^{2} + (5.47 u)^{3} + (6.71 u)^{4} \right]^{-1/4}
\end{align}
We can then write $\sigma_R^2$ as
\begin{align}
\label{sigma squared app}
\sigma_{R}^{2} = \left(\frac{H^{2}}{\dot{\phi}_{0}}\frac{2}{5}\frac{1}{\Omega_{m}H_{0}^{2}R^2}\right)^{2}C_{2}(\mu,m){\cal J}
\end{align}
where
\begin{align}
{\cal J} = \frac{1}{2\pi^{2}}\int\limits_{0}^{\infty}dx x^{3}T(x/R)^{2}W(x)^2
\end{align}
and $W(x)\equiv W_R(x/R)$ is independent of $R$.

The Fourier transform of the halo two-point gives the halo power spectrum
\begin{align}
P_{hh}({ q}) = \int d^{3}{\bf x}\left<\delta_{h}({\bf x})\delta_{h}({\bf 0})\right>e^{-i{\bf q}\cdot {\bf x}}.
\end{align}
Fourier transforming (\ref{2 point bias expan1}) and plugging in (\ref{delta to xi main text}) to express the correlation functions of $\delta_{R}({\bf k})$ in terms of those of $\zeta_{\bf k}$, we find for $q<<1/R$:
\begin{align}
\label{halo power bias expan}
P_{hh}({\bf q}) &= b_{1}^{2}\alpha_{R}(q)^{2}P_{\zeta}({\bf q}) + 2b_{1}b_{2}\alpha_{R}({\bf q})\int\frac{d^{3}k}{(2\pi)^{3}}\alpha_{R}(k)^2B_{\zeta}({\bf q},{\bf k},-{\bf k}-{\bf q})\cr 
&\ \ \ \ \ \ \ \ + b_{2}^{2}\int\frac{d^{3}k_{1}}{(2\pi)^{3}}\frac{d^{3}k_{2}}{(2\pi)^{3}}\alpha_{R}(k_{1})^{2}\alpha_{R}(k_{2})^{2}N_{\zeta}^{(4)}({\bf k}_{1},{\bf q}-{\bf k}_{1},{\bf k}_{2},-{\bf k}_{2}-{\bf q}).
\end{align}
To condense the expression we have defined 
\begin{align}
\alpha_{R}(k) =\frac{2k^{2}}{5\Omega_{m}H_{0}^{2}}T(k)W_{R}(k).
\end{align}

The wave-vectors integrated over in the integrals of (\ref{halo power bias expan}) are of order $1/R$.  Since we are interested in $q << 1/R$ the curvature bispectrum and trispectrum appearing in (\ref{halo power bias expan}) are in their squeezed and collapsed configurations.  Equations (\ref{B squeezed}) and (\ref{analytic 4 point comp}) imply the strongest small $q$ scaling of the primordial squeezed bispectrum and collapsed trispectrum are $1/q^{3-\alpha_{-}}$ and $1/q^{3-2\alpha_{-}}$.  Note that the bispectrum's contribution to the halo power spectrum is suppressed by a factor of $\alpha_R(q)\propto q^2$, so that term goes like $1/q^{1-\alpha_-}$.

\begin{figure}
\includegraphics[width=2in]{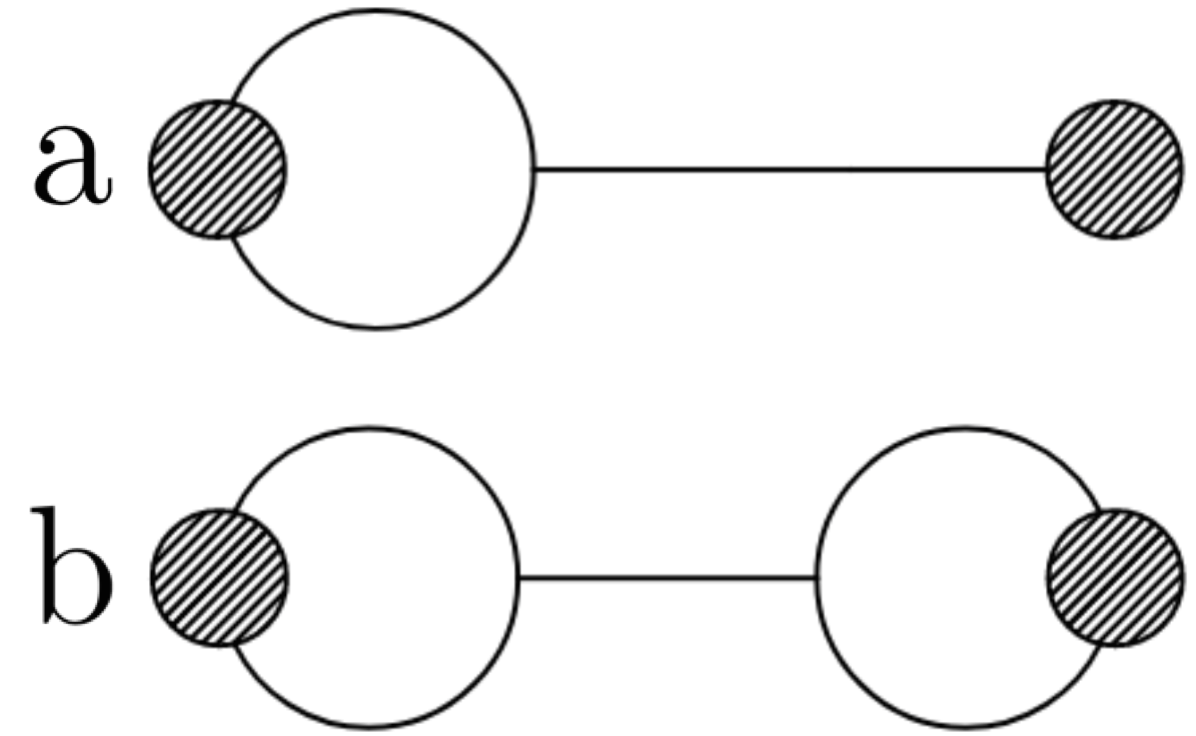}
\caption{A diagrammatic representation of terms contributing to the galactic halo power spectrum.  Cf. Fig. {\ref{fig:feynman 2 pt}}.}
\label{fig:schematic 2 pt}
\end{figure}

\begin{figure}
\includegraphics[width=5in]{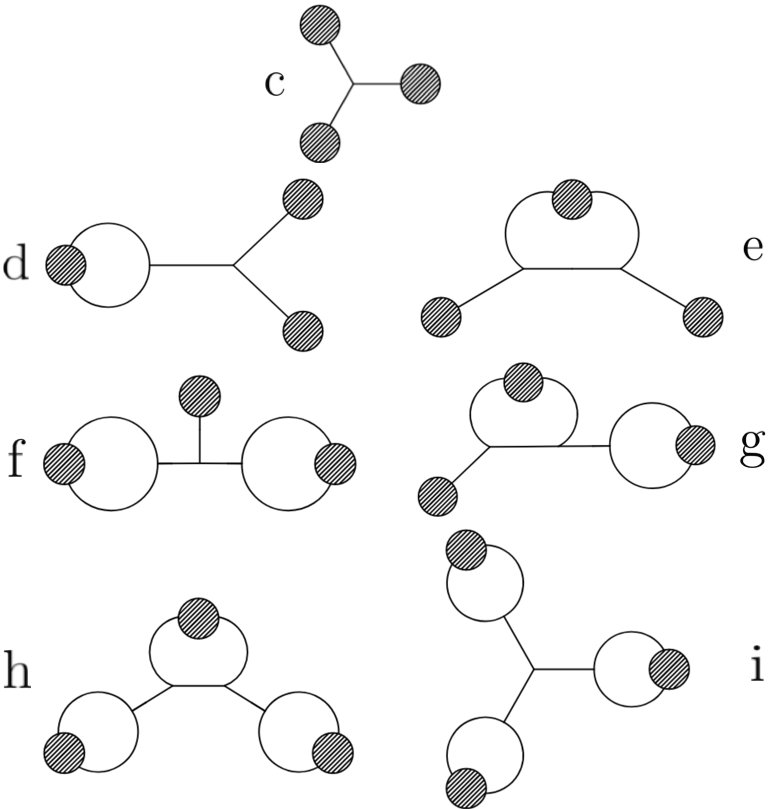}
\caption{A diagrammatic representation of terms contributing to the galactic halo bispectrum.  Cf. Fig. {\ref{fig:feynman 3 pt}}.}
\label{fig:schematic 3 pt}
\end{figure}

An intuitive picture of the non-Gaussian contributions to (\ref{halo power bias expan}) is given by Fig. \ref{fig:schematic 2 pt}.  The shaded circles represent the halo overdensity, while the lines they are attached to are $\zeta$ legs. In these graphs, the external $\zeta$ legs are each multiplied by $\alpha_R$.  If one $\zeta$ leg is attached to a shaded circle it carries a soft wave-vector and a factor of $b_{1}$.  If two legs are attached to a shaded circle, they carry equal and opposite wave-vectors with magnitude approximately $1/R$.  In this case, the shaded circle also contains a factor of $b_{2}$ and a wave-vector integral. 

The halo power spectrum is then\footnote{In writing (\ref{galactic power spectrum}) we have used $\int\limits_{0}^\infty dx x^{3-n\alpha_{-}}T(x/R)^{2}W(x)^{2} \simeq \int\limits_{0}^\infty dx x^{3}T(x/R)^{2}W(x)^{2}$, where $n$ is an O(1) integer.}
\begin{align}\label{galactic power spectrum}
P_{hh}(q) &= P_{hh}^{G}(q)\left[1+\gamma(\mu,m) \left(2\frac{\beta(\mu,m)}{(qR)^{2-\alpha_-}T(q)}+\frac{\beta(\mu,m)^2}{(qR)^{4-2\alpha_-}T(q)^2}\right)\right]
\end{align}
where 
\begin{align}
&P_{hh}^G(q)=b_1^2 P_{mm}(q)\cr
&P_{mm}(q)=R^3C_2(\mu,m)\left(\frac{H^2}{\dot \phi_0}\right)^2\left(\frac{2}{5 \Omega_m H_0^2 R^2}\right)^2(qR) T(q)^2\cr
&\gamma(\mu,m)=\frac{9\mu^2H^2}{(\mu^2+m^2)^2+9\mu^2H^2}\cr 
&\beta(\mu,m)=\frac{6}{5}\frac{b_2}{b_1}\frac{H^2}{\dot \phi_0}\frac{2}{5\Omega_m H_0^2 R^2}{\cal J}f_{NL}\sqrt{2\pi^{2}\Delta^{2}_{\zeta}}C_{2}(\mu,m)^{\frac{3}{2}}\frac{(\mu^2+m^2)^2}{(3\mu/2)^2H^2}.
\end{align}
$P_{mm}$ denotes the ``matter-matter'' power spectrum, \textit{i.e.}, the Fourier transform of $\langle \delta_R({\bf x})\delta_R({\bf y})\rangle$.

Since $0<\gamma(\mu,m)<1$, it is simple to show that $P_{hh}(q)$ is positive definite, as it must be.  Note that for $f_{NL}<0$, this would not be true at very small wave-vectors without the contribution due to the four-point function of $\zeta$.  The scale non-Gaussianities begin to dominate is $(qR)^2\sim \beta(\mu,m)\propto f_{NL}$ (up to $(qR)^{\alpha_-}$ terms).  Current measurements of the galactic power spectrum have not seen significant deviations from Gaussian initial conditions at wave-vectors around $q \sim h/(100 \ {\rm Mpc})$~\cite{Beutler}. 

In the threshold model, we find that $\beta\propto R^2$, indicating that the scale at which non-Gaussianities begin to dominate is independent of model parameter $R$.  

On the other hand, we can also compute the matter-halo cross correlation power spectrum $P_{hm}(q)$, which corresponds to the two-point function $\langle \delta_h({\bf x})\delta_R({\bf y})\rangle$.  The ``$h$" in $P_{hm}$ stands for halo, and the ``$m$" for matter. The result is
\begin{align}
P_{hm}(q)=b(q)P_{mm}(q)
\end{align}
where
\begin{align}
b(q)\equiv b_1+b_1\gamma(\mu,m)\beta(\mu,m)\frac{1}{(qR)^{2-\alpha_-}T(q)}.
\end{align}
This implies a scale-dependent bias:\footnote{Recall that we have neglected the time evolution of the distribution of galaxies after they have formed.}
\begin{align}
\Delta b(q)=b_1\gamma(\mu,m)\beta(\mu,m)\frac{1}{(qR)^{2-\alpha_-}T(q)}.
\end{align}
Note that $P_{hh}$ can be written in this notation as:
\begin{align}
P_{hh}(q)=\left(b(q)^2+b_1^2\beta(\mu,m)^2\gamma(\mu,m)\left(1-\gamma(\mu,m)\right)\frac{1}{(qR)^{4-2\alpha_-}T(q)^2}\right)P_{mm}(q).
\end{align}
In this form, the second term in the brackets is due to stochastic bias.  Note that this term is proportional to $1-\gamma(\mu,m)$, which approaches $0$ in the limit that $\mu\gtrsim m$ as $\mu/H$ and $m/H$ go to zero.  This suppression is evident in Fig. \ref{fig:alpha2pt}.  If the stochastic bias were zero, then the purple curves' minimum value would be $0$.  Since they all reach a minimum value less than around $0.1$, this indicates that the stochastic bias is small in the $\mu\sim m$ regime. However, for $\mu<<m$ the stochastic bias can become large, see Fig. \ref{fig:mum2pt}.  As we will show toward the end of this section, for $\mu$ several orders of magnitude smaller than $m$, other contributions to the power spectrum that we have neglected become important.

In figures \ref{fig:alpha2pt} and \ref{fig:mum2pt}, we plot the ratio of the galactic halo power spectrum in quasi-single field inflation divided by the Gaussian contribution $P_{hh}^G$.  Notice that for reasonable model parameters, $P_{hh}(q)$ begins to differ from $P_{hh}^G(q)$ at $q\sim 0.005 h/{\rm Mpc}$.  The difference becomes very large for values of $q$ significantly less than this.  Figures \ref{fig:alpha2pt} and \ref{fig:mum2pt} use $f_{NL}=\pm 10$, and various values for $\alpha_-$ and $\mu$.  

\begin{figure*}[tp]
\centering
\includegraphics[width=5in]{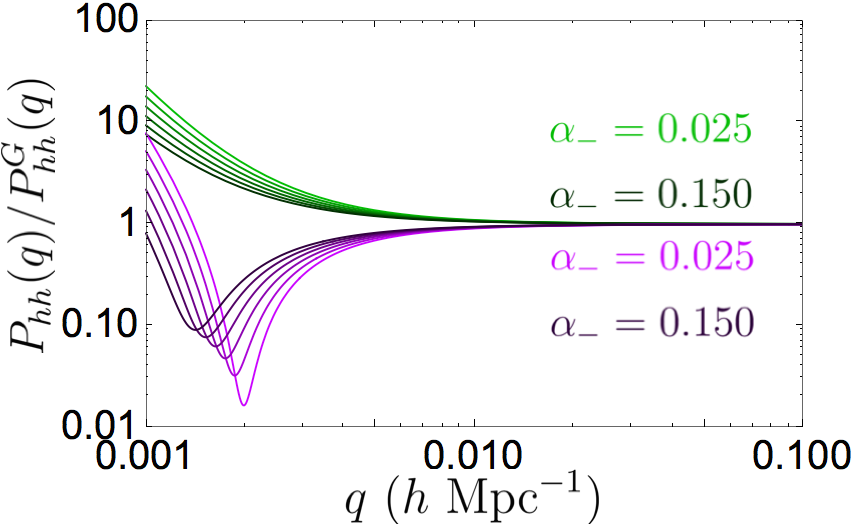}
\caption{We plot the ratio of the galactic halo power spectrum in quasi-single field inflation to the halo power spectrum in which there are no primordial non-Gaussianities for a range of $\alpha_{-}$: $\alpha_-=0.025$ (lightest), $0.050$, $0.075$, $0.100$, $0.125$, $0.150$ (darkest).  We plot for $\mu=m$ and $f_{NL}=10$ (green) and $f_{NL}=-10$ (purple).}  
\label{fig:alpha2pt}
\end{figure*}

\begin{figure*}[tp]
\centering
\includegraphics[width=5in]{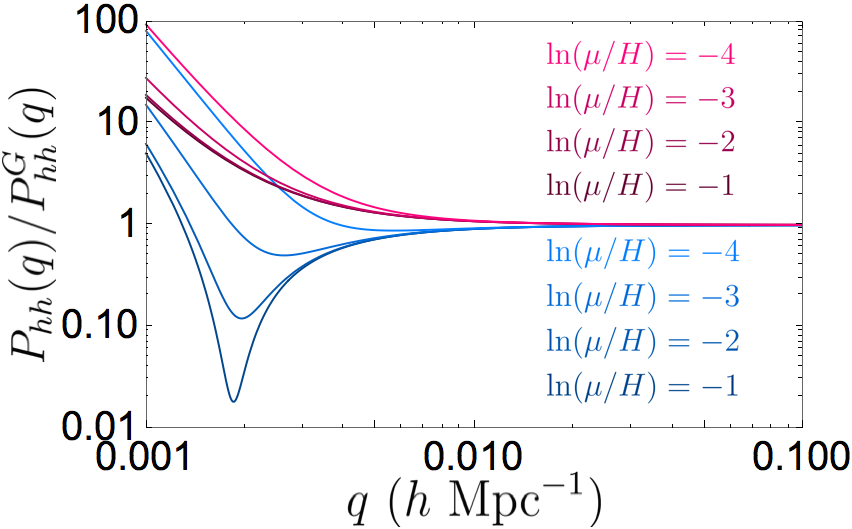}
\caption{We plot the ratio of the galactic halo power spectrum in quasi-single field inflation to the halo power spectrum in which there are no primordial non-Gaussianities for a range of $\mu$. We plot for $\ln(\mu/H)=-1$ (darkest), $-2$, $-3$, $-4$ (lightest), with $\alpha_-=0.05$ and $f_{NL}=10$ (pink) and $f_{NL}=-10$ (blue).}
\label{fig:mum2pt}
\end{figure*}

Let us now study the halo three-point function given in equation (\ref{biashalo3}). 
The non-Gaussian contributions are depicted in Figure \ref{fig:schematic 3 pt}.  Fourier transforming equation (\ref{biashalo3}), we find that the bispectrum of the halo overdensity is 
\begin{align}
&B_{hhh}({\bf q}_1,{\bf q}_2,{\bf q}_3)=b_1^3 \alpha_R(q)^3 B_\zeta({\bf q}_1,{\bf q}_2,{\bf q}_3)\cr 
&+\left[2b_1^2b_2\alpha_R(q_2)^2\alpha_R(q_3)^2P_\zeta(q_2)P_\zeta(q_3)+b_1^2b_2\alpha_R(q_2)\alpha_R(q_3)\int \frac{d^3 k}{(2\pi)^3}\alpha_R(k)^2N_\zeta^{(4)}({\bf k},{\bf q}_1-{\bf k},{\bf q}_2, {\bf q}_3)\right.\cr 
&\left.+b_1 b_2^2 \alpha_R(q_3)\int\frac{d^3k_1}{(2\pi)^3}\frac{d^3 k_2}{(2\pi)^3}\alpha_R(k_1)^2\alpha_R(k_2)^2 N_\zeta^{(5)}({\bf k}_1,{\bf q}_1-{\bf k}_1,{\bf k}_2,{\bf q}_2-{\bf k}_2,{\bf q}_3)+\text{cyc. perm}({\bf q}_1,{\bf q}_2,{\bf q}_3)\right] \cr 
&+b_2^3  \int \frac{d^3k_1}{(2\pi)^3}\frac{d^3 k_2}{(2\pi)^3}\frac{d^3 k_3}{(2\pi)^3}\alpha_R(k_1)^2\alpha_R(k_2)^2\alpha_R(k_3)^2N_\zeta^{(6)}({\bf k}_1,{\bf q}_1-{\bf k}_1,{\bf k}_2,{\bf q}_2-{\bf k}_2,{\bf k}_3,{\bf q}_3-{\bf k}_3).\cr
\end{align}

Similar to the calculation of the two-point, we can simplify the wave-vector integrals to express the bispectrum as
\begin{align}
\label{full bispectrum eqt}
&B_{hhh}({\bf q}_1,{\bf q}_2,{\bf q}_3)=2b_1^2b_2R^6\left(\frac{H^2}{\dot \phi_0}\right)^4\left(\frac{2}{5\Omega_m H_0^2 R^2}\right)^4C_2^2\Bigg[T(q_1)^2T(q_2)^2q_1 q_2 R^2.\cr 
&+\left.\omega(\mu,m)\Bigg(\beta(\mu,m)\frac{q_1^2}{q_2q_3}T(q_1)T(q_2)T(q_3)\right.\cr
& +\beta(\mu,m)^2T(q_2)T(q_3)(qR)^{\alpha_{-}}\bigg[\frac{q_2^2}{R^2q_1^3 q_3}+\frac{q_3^2}{R^2q_1^3 q_2} + 2\big(1+\frac{1}{2}(q R)^{\alpha_-}\big)\frac{1}{R^2q_2 q_3}\bigg]\cr
&+\beta(\mu,m)^3 T(q_3)(qR)^{2\alpha_{-}}\bigg[\frac{q_3^2}{q_1^3 q_2^3 R^4}+ 2\big(1-\frac{1}{12}\left(q R\big)^{\alpha_-}\right)\frac{1}{R^4 q_2^3 q_3}+2\big(1-\frac{1}{12}\left(q R\big)^{\alpha_-}\right)\frac{1}{R^4 q_1^3 q_3}\bigg]\cr &+\beta(\mu,m)^4(qR)^{3\alpha_{-}}\left(2+\left(q R\right)^{\alpha_-}\right)\frac{1}{R^6 q_1^3 q_2^3}\Bigg)\Bigg]+\text{ cyc. perm}(q_1,q_2,q_3).
\end{align}
where $q \equiv {\rm max}(q_{i})$, and
\begin{align}
\omega(\mu,m)=\frac{b_1^2}{4b_2^2}\frac{1}{{\cal J}C_2}\left(\frac{\dot \phi_0}{H^2}\right)^2\left(\frac{5\Omega_m H_0^2R^2}{2}\right)^2\gamma(\mu,m).
\end{align}
Again, the scale at which the non-Gaussian contributions begin to dominate is $(qR)^2\sim\beta(\mu,m)$, which means the galactic power spectrum and bispectrum both begin to deviate from their Gaussian contributions at roughly the same scale.  Since it is easier to measure the halo two-point function than the halo three-point function, it is more likely that we will see these non-Gaussian effects in the halo two-point before we see them in the three-point. 

The equilateral configuration of the galactic halo bispectrum is plotted in Figs. \ref{fig:alpha3pt} and \ref{fig:mum3pt} for various values of $\alpha_-$ and $\mu$.  Note that we have scaled the bispectrum by its value when $V'''=0$,
\begin{align}
B_{hhh}^G({\bf q}_1,{\bf q}_2,{\bf q}_3)=2b_1^2b_2R^6\left(\frac{H^2}{\dot \phi_0}\right)^4\left(\frac{2}{5\Omega_m H_0^2 R^2}\right)^4C_2^2T(q_1)^2T(q_2)^2q_1 q_2 R^2 + {\rm ~cyc.~ perm}(q_1,q_2,q_3).
\end{align}
In the equilateral configuration with $f_{NL}<0$, this scaled bispectrum never falls significantly below unity.  Note also that it rises more rapidly than the scaled power spectrum shown in Figs. \ref{fig:alpha2pt} and \ref{fig:mum2pt} as $q$ becomes small.

Equation (\ref{full bispectrum eqt}) expresses the bispectrum in terms of the magnitude of the wave-vectors ${\bf q}_{1}, {\bf q}_{2}$ and ${\bf q}_{3}$.  It could also be expressed in terms of $q_{1}$ and $ q_{2}$ and the angle between them. This angular dependence is usually displayed as a multipole expansion. 


\begin{figure*}[tp]
\centering
\includegraphics[width=5in]{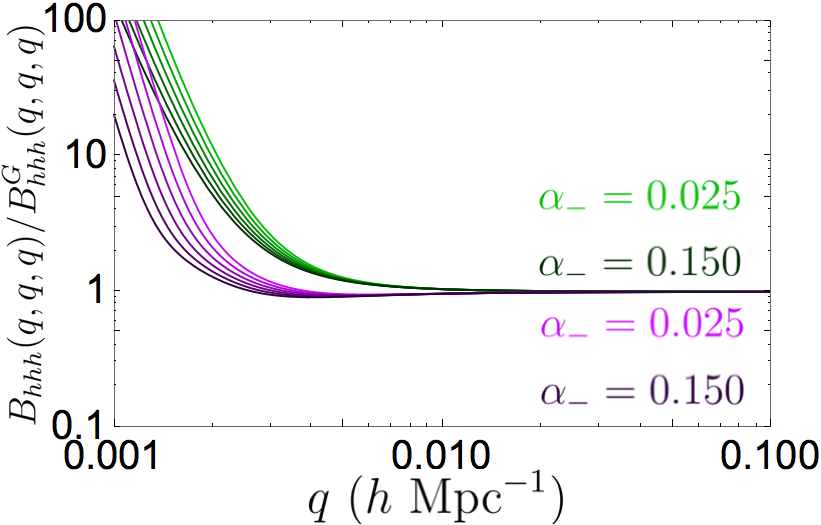}
\caption{We plot the ratio of the galactic halo bispectrum in quasi-single field inflation to the galactic halo bispectrum with no primordial non-Gaussianities ($B_{hhh}^G$) in the equilateral configuration for a range of $\alpha_{-}$:  $\alpha_{-} = 0.025 $ (lightest), $0.050$, $0.075$, $0.100$, $0.125$, $0.150$ (darkest).  We plot for $\mu=m$ and  $f_{NL}=10$ (green) and $f_{NL}=-10$ (purple).}
\label{fig:alpha3pt}
\end{figure*}

\begin{figure*}[tp]
\centering
\includegraphics[width=5in]{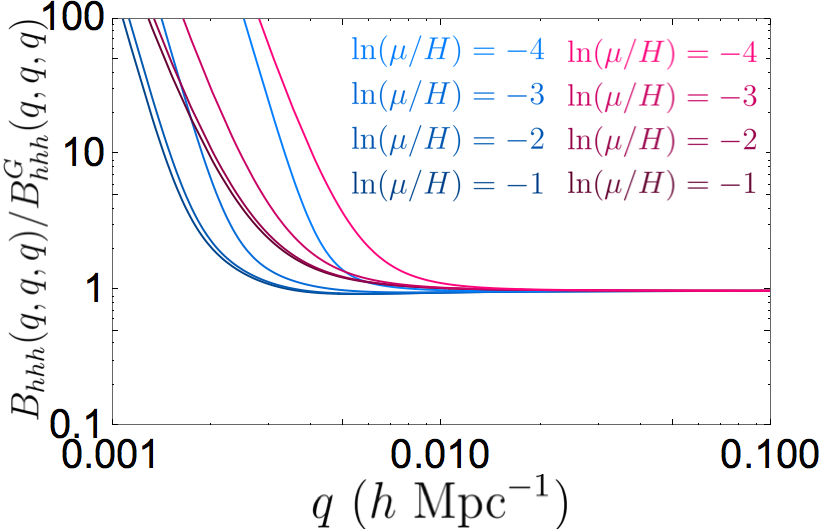}
\caption{We plot the ratio of the galactic halo bispectrum in quasi-single field inflation to the galactic halo bispectrum with no primordial non-Gaussianities ($B_{hhh}^G$) in the equilateral configuration for a range of $\mu$: $\ln(\mu/H)=-1$ (darkest), $-2$, $-3$, $-4$ (lightest).  We plot for $\alpha_-=0.05$ and $f_{NL}=10$ (pink) and $f_{NL}=-10$ (blue).}
\label{fig:mum3pt}
\end{figure*}

Currently, there are measurements of the galaxy bispectrum at wave-vectors as small as about $h/(20 ~{\rm Mpc})$~\cite{Gil Martin}. There is no evidence in this data for the type of effects we have found. 

 We have ignored the evolution of the galactic halo distribution after their collapse. These effects are $O(1)$.  However, we do not expect that including them  greatly shifts at what scale non-Gaussianities or their rapid growth become important. One can include these effects either by numerical simulation or analytic methods~\cite{Fry:1983cj,Mirbabayi:2014zca,Angulo:2015eqa}. Evolution during this period is expected to decrease the influence of bias, drawing the galactic distribution closer to the dark matter distribution.  Some of these effects cancel out in the ratios we have plotted.  
 
We have chosen to plot the power spectrum and bispectrum scaled by $P_{hh}^G$ and $B_{hhh}^G$ since these ratios are less sensitive to the value of $R$ than the power spectrum and bispectrum alone.

\begin{figure*}[tp]
\centering
\includegraphics[width=2.5in]{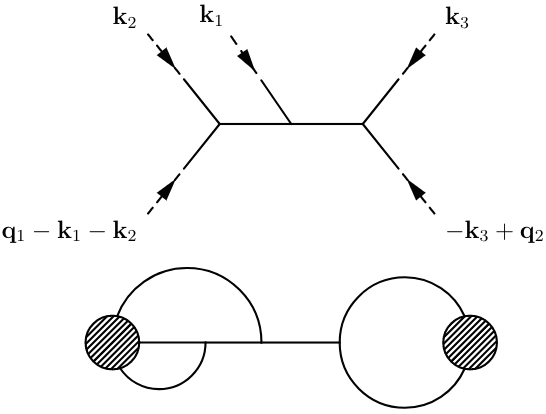}
\caption{The above diagram can contribute significantly to the galactic halo power-spectrum if $|V'''|/H$ is not very small.  However, it can be ignored as long as $|V'''|/H<<1$.  In the context of eq. (\ref{n point power counting}), this is a $p=1$, $j=0$ term.}
\label{fig:extradiagram}
\end{figure*}

It is possible to use the methods developed here to consider even higher correlations of the halo overdensity. 
The dependence of galactic halo $n$-point correlations on the parameters $V'''$, $q$, and $R$ in quasi-single field inflation with the halo number density modeled by eq. (\ref{halo formation model}) is given by
\begin{align}\label{n point power counting}
\langle\delta_h^n\rangle\sim R^{3(n-1)}(qR)^{n-1}\left[1+\sum_{i=n-2}^{2n-2}\left(\frac{V'''}{H(qR)^2}\right)^i \ \sum_{j=0}^{n-1}\sum_{p=0}^\infty(qR)^{3j}\left(\frac{V'''}{H}\right)^p\right]
\end{align}
where for simplicity, factors of $(qR)^{\alpha_-}$ have been set to unity.  In our analysis of the power spectrum ($n=2$) and the bispectrum ($n=3$), we have included only the $j=p=0$ terms in the sums.  

Recall, the validity of our calculations relies on the several assumptions.  First of all we have assumed that $\alpha_-=(\mu^2+m^2)/3H^2<<1$.  However, we must also have $\alpha_-\gtrsim 1/60$ or else superhorizon evolution would have persisted to the end of inflation. Finally, we assumed $qR<<1$ and $|V'''|/H<<1$.  Note that for fixed $|f_{NL}|=10$ and $\alpha_-=0.05$, then $|V'''|/H> 1$ for $\mu<0.005$.  Therefore, our results do not apply at very small $\mu/m$.  For $|V'''|/H$ not small, we would need to include additional contributions, \textit{e.g.}, the diagram shown in Fig. ~\ref{fig:extradiagram}.

\section{Conclusions}
The $1/q^3$ dependence of the de-Sitter propagator for massless scalar fields implies that if the primordial curvature fluctuations are non-Gaussian, they have the potential to give rise to enhancements in the correlations of  biased objects at small wave-vectors~\cite{AGW,Dalal:2007cu}. This effect cannot be produced by nonlinear gravitational evolution without primordial non-Gaussianities. The main goal of this paper was to explore these enhancements within quasi-single field inflation.

We developed a method to analytically compute the correlation functions of the curvature perturbation $\zeta$ in quasi-single field inflation in the limit of small $m/H$ and $\mu/H$.  We computed the three- and four-point functions of $\zeta$ for arbitrary external wave-vectors and computed the five- and six-point functions in the kinematic limits that give the strongest long wavelength enhanced contributions to the three-point function of the galactic halo overdensity $\delta_{h}$. 

We applied these results to the computation of the two- and three-point correlations of $\delta_{h}$  ({\it i.e.}, the power spectrum and bispectrum).  For model parameters consistent with the constraints on $f_{NL}$, we found that non-Gaussian contributions to these correlation functions are larger than the Gaussian ones at scales around $ h/(200 {\rm Mpc})$. Even larger scales will be probed in upcoming large scale surveys such as SPHEREx. Prospects for future improvements in measurements of the galactic power spectrum and bispectrum are reviewed in~\cite{Alvarez:2014vva}.

After making a number of approximations, we  obtained analytic expressions for the power spectrum and bispectrum\footnote{Since galactic halos are biased objects, even if the primordial fluctuations are Gaussian a halo bispectrum is not zero.}  of $\delta_{h}$ that are valid at small wave-vectors.  We studied the dependence of the stochastic bias on the parameters $\mu$ and $m$, and found that it could be small or significant depending on the values of $\mu$ and $m$.

The departure from the predictions of Gaussian primordial perturbations in both the equilateral configuration of the bispectrum and the power spectrum begin at wave-vectors around $h/(200 {\rm Mpc})$ (when $|f_{NL}|$ is near its upper bound). However, for the bispectrum the deviation grows much more rapidly as the wave-vectors decrease than in the power spectrum.  Unfortunately, it is more difficult to measure the three-point correlation than the two-point correlation of $\delta_{h}$.  If these enhancements exist, it is more likely we will first see them in the power spectrum than in the bispectrum.  Finally, we identified the scaling of the $n$-point function of $\delta_{h}$. 

The calculations (at small wave-vectors) of the galactic power spectrum and bispectrum presented in this paper can be improved and made more model independent. We hope to address this in future work.

\section*{Acknowledgements}
We would like to thank Olivier Dor\'e, Roland de-Putter, Daniel Green and Mikhail Solon for useful discussions.  This work was supported by the DOE Grant DE-SC0011632. We are also grateful for the support provided by the Walter
Burke Institute for Theoretical Physics.

\appendix

\section{Numerical Checks}
\label{numerical checks appendix}
In this appendix we check that the analytical results we derived for the two- and three-point functions of $\zeta$ agree with the numerical evaluation of these quantities. First, consider the two-point function.  In equation (\ref{2pt:pipi}), we absorbed all of the $\mu$ and $m$ dependence of the curvature perturbation power spectrum into the constant $C_{2}(\mu, m)$.  We can express this quantity in terms of the exact mode functions of $\pi$ as
\begin{align}
\label{2 point exact app}
C_{2}(\mu, m) = \sum\limits_{i}|\pi^{(i)}(0)|^{2}.
\end{align}
We found in equation (\ref{leading C2}) that the  leading behavior was
\begin{align}
\label{2 point approx app}
C_{2}(\mu,m) \simeq \frac{1}{2} + \frac{9\mu^{2}H^{2}}{2(\mu^{2} + m^{2})^{2}}.
\end{align}
up to terms suppressed by $\alpha_{-}$.  By extending the numerical techniques developed in \cite{Assassi:2013gxa} and \cite{An:2017hlx} to the region of small $\mu/H$ and $m/H$ we can compute (\ref{2 point exact app}) numerically.  In Fig. \ref{fig:2point} we compare (\ref{2 point approx app}) to the numerical evaluation of (\ref{2 point exact app}).  The fit is good even for modest values of $\mu/H$ as long as $m/H$ is small.

\begin{figure*}[tp]
\centering
\includegraphics[width=3.99in]{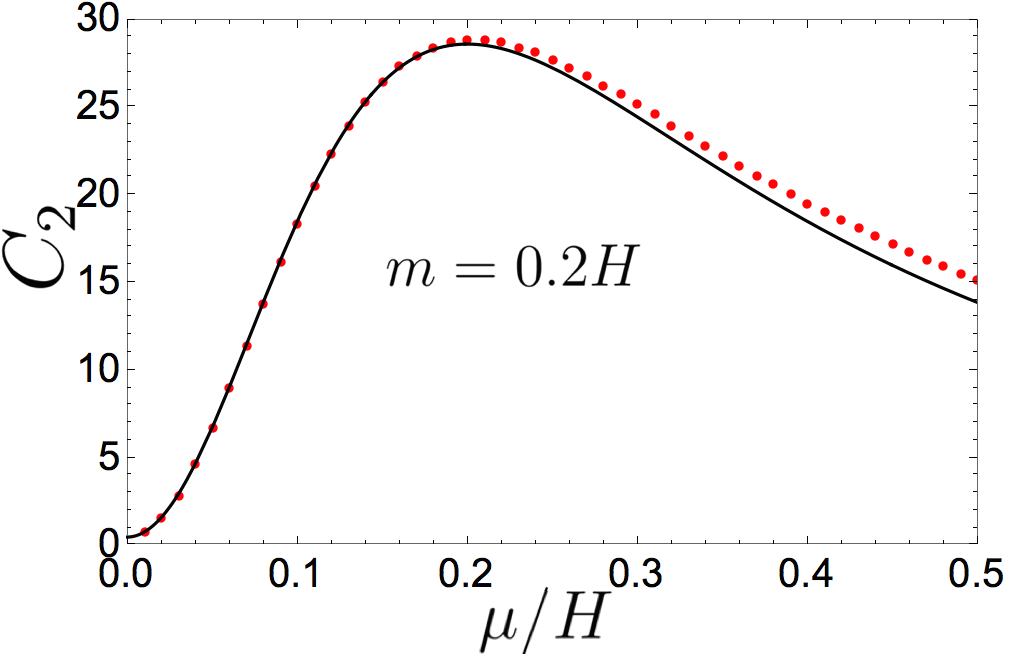}
\includegraphics[width=3.99in]{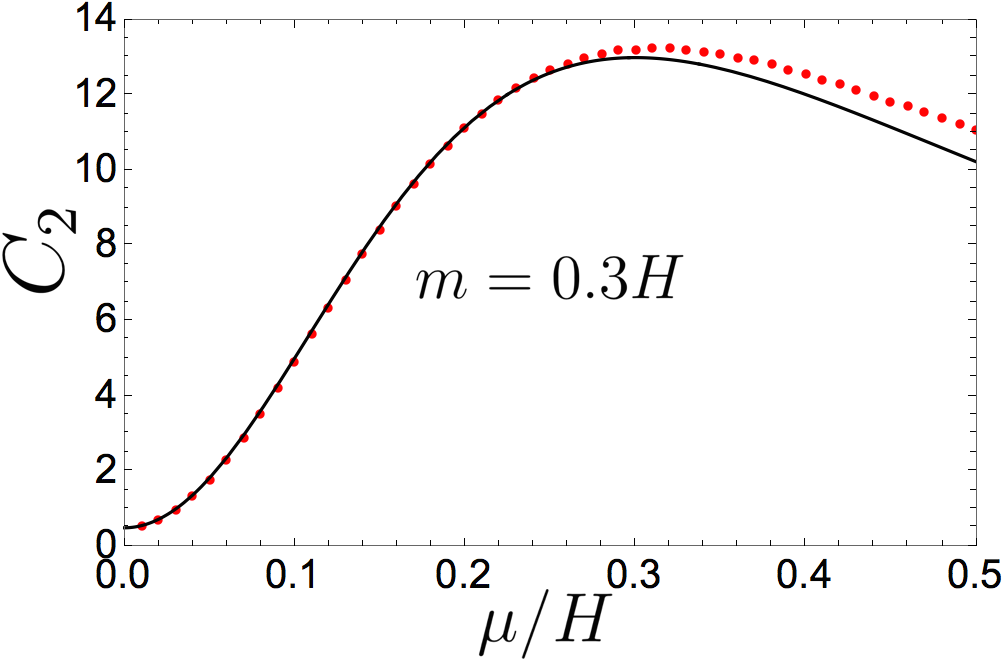}
\caption{We compare the power spectrum (\ref{power spectrum}) (red) computed with the numeric mode functions against the leading $\mu$ and $m$ expression (\ref{leading C2}) (black) for $m = 0.2H$ (top) and $m = 0.3H$ (bottom).}
\label{fig:2point}
\end{figure*}

\begin{figure*}
\centering
\includegraphics[width=4in]{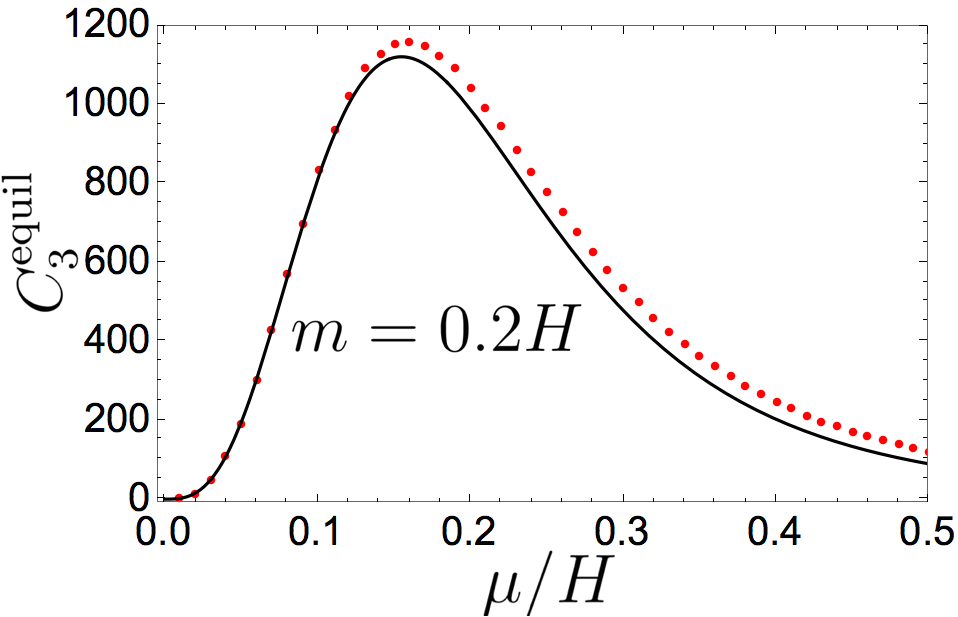}
\includegraphics[width=4in]{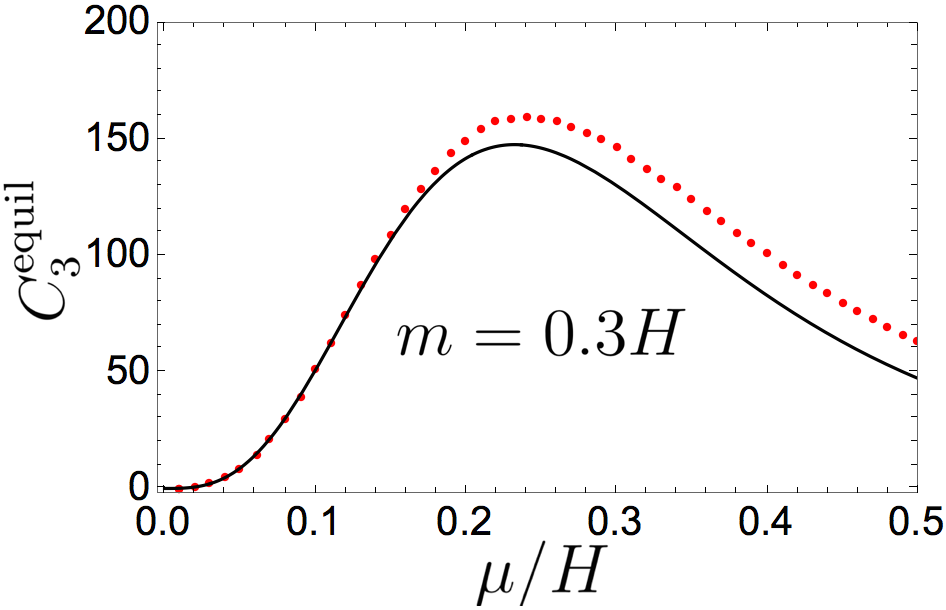}
\caption{We plot the numerical evaluation of (\ref{3 point equil integral appendix}) (red) with the leading $\mu$ and $m$ expression (\ref{copy formula leading 3 pt equil appendix}) (black) for $m = 0.2H$ (top) and $m = 0.3H$ (bottom).}
\label{threepointequil}
\end{figure*}

\begin{figure*}
\centering
\includegraphics[width=4in]{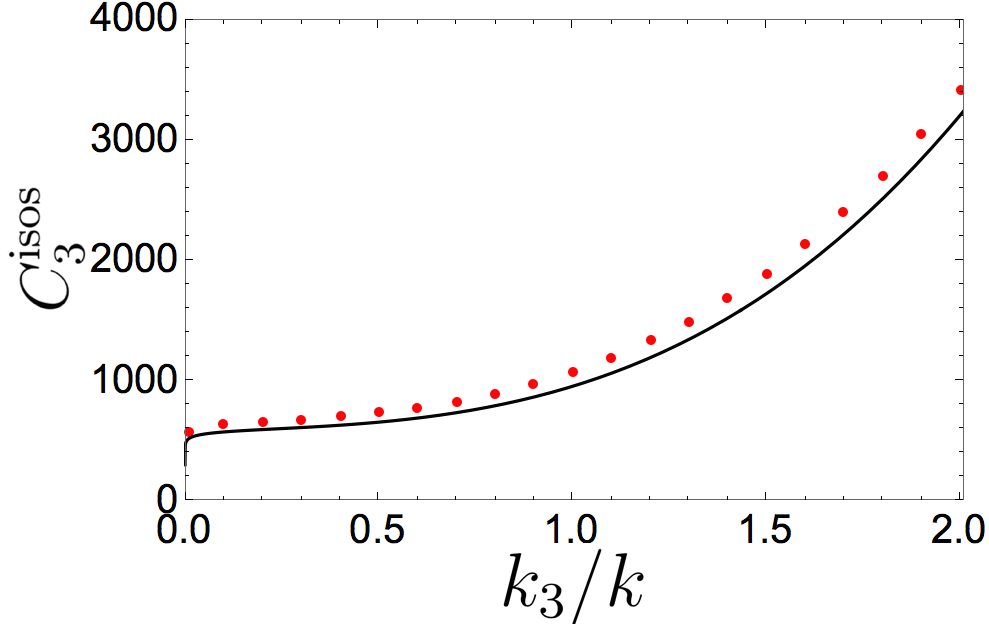}
\caption{We plot the numerical evaluation of (\ref{3 point master integral}) (red) taking $\mu = 0.3H$, $m = 0.2H$ against (\ref{C3 isos}) (black).}
\label{3pointisos}
\end{figure*}

To determine the accuracy of our formula for the bispectrum of $\zeta$ (\ref{3 point leading behavior}), we compare it with the numerical evaluation of the exact result (\ref{3 point}) in a couple of kinematic limits.  Let's first consider the equilateral configuration.  We define $C_{3}^{\rm equil}(\mu, m)$ to be the integral in eq. (\ref{3 point}) in the equilateral configuration:
\begin{align}
\label{3 point equil integral appendix}
C^{{\rm equil}}_{3}(\mu,m) \equiv \int\limits_{-\infty}^{0}\frac{d\eta}{\eta^{4}}{\rm Im}\left[\left(\pi^{(1)}(0)s^{(1)*}(\eta) + \pi^{(2)}(0)s^{(2)*}(\eta)\right)^{3}\right].
\end{align}
Equation (\ref{3 point equil B}) gives the leading behavior of this integral for small $\alpha_-$, which we reproduce here for convenience
\begin{align}
\label{copy formula leading 3 pt equil appendix}
C^{{\rm equil}}_{3}(\mu,m) = \frac{3(3\mu/2)^{3}H^5}{2\left(\mu^{2} + m^{2}\right)^{4}}.
\end{align}
Again, we can use the same numerical techniques to compute (\ref{3 point equil integral appendix}).  However, there is a subtlety in its evaluation that needs to be addressed.  

As mentioned in section \ref{three point function section}, the integral is naively IR divergent because of the factor of $1/\eta^{4}$.  However, through the commutation relations, we proved the leading IR behavior is $(-\eta)^{-1 + 2\alpha_{-}}$ in the IR, and that the integral is IR finite.  However, numerical error prevents the coefficients in front of the potentially IR divergent terms from canceling exactly, giving rise to spurious infinities.  The way around this is to define the integrand piecewise about some point $\eta_{IR}$.  For $\eta < \eta_{IR}$ we use the numerical mode functions in the integrand, and for $\eta > \eta_{IR}$ we set the integrand equal to $a(-\eta)^{-1 + 2\alpha_{-}}$, where $a$ is some proportionality constant that can be obtained by fitting the integrand to the correct power law.   
 
In Fig. \ref{threepointequil} we compare (\ref{copy formula leading 3 pt equil appendix}) to the numerical evaluation of (\ref{3 point equil integral appendix}).  As expected the fit is better for smaller values of $m$, however it is still accurate to around $25$ percent even for $\mu = 0.5H$ and $m = 0.3H$. 

The previous tests have confirmed the $\mu$ and $m$ dependence of our analytic expressions.  To test the dependence on the external wave-vectors, we consider the isosceles configuration in which $k_1=k_2 \equiv k$ and $0 \le k_3 \le 2k$.  In this limit, equation (\ref{3 point leading behavior}) becomes
\begin{align}
B^{\rm isos}_{\zeta}(k, k_{3}) = -\left(\frac{H^{2}}{\dot{\phi_{0}}}\right)^{3}\left(\frac{V'''}{H}\right)\frac{1}{k^{3}k_{3}^{3}}C_3^{\rm isos}
\end{align}
where we have defined
\begin{align}\label{numerical isos 3 config lab}
C_3^{\rm isos}\equiv2 \ {\rm Im}\int\limits_{-\infty}^{0}\frac{d\eta}{\eta^{4}}\left(\sum\limits_{i}\pi^{(i)}(0)s^{(i)*}(\eta)\right)^{2}\left(\sum\limits_{i}\pi^{(i)}(0)s^{(i)*}\left(\frac{k_{3}}{k}\eta\right)\right).
\end{align}
Equation (\ref{C3 equil IR approx}) approximates $C_{3}^{\rm isos}$ as
\begin{align}
\label{C3 isos}
C_{3}^{\rm isos}(\mu,m,k,k_{3}) \simeq \frac{(3\mu/2)^{3}}{\left(\mu^{2} + m^{2}\right)^{4}}\left[2\left(\frac{k_{3}}{k}\right)^{\alpha_{-}} + \frac{k_{3}^{3}}{k^3}\right].
\end{align}
In Fig. \ref{3pointisos} we plot (\ref{C3 isos}) against the numerical evaluation of (\ref{numerical isos 3 config lab}). The errors are around 10 percent for each data point, suggesting the error is not in the wave-vector dependence of the formula, but rather in its $\mu$ and $m$ dependent normalization.

\section{Outline of Five- and Six-Point Calculations}
\label{appendix 5 and 6}
In order to compute the three-point correlation of biased objects, it is necessary to compute five- and six-point correlation functions in certain kinematic regimes. The contribution due to the diagram in panel $f$ of Fig.~\ref{fig:feynman 3 pt},  $N_{\zeta,f}^{(5)}$, can be computed using the commutator form of the in-in formalism:
\begin{align}
N_{\zeta,f}^{(5)}&({\bf k}_1,{\bf q}_1-{\bf k}_1,{\bf k}_2,{\bf q}_2-{\bf k}_2,{\bf q}_3) =-\left(\frac{H^2}{\dot \phi_0}\right)^2\left(\frac{V'''}{H}\right)^3\frac{8}{k_1^6 k_2^6 q_1^3 q_2^3 q_3^3}(q_1 q_2)^{\alpha_-}\cr 
&\int_{-\infty}^0\frac{d\tau}{\tau^{4}}\int_{-\infty}^\tau \frac{d\tau'}{\tau'^{4}}\int_{-\infty}^{\tau'}\frac{d\tau''}{\tau''^4}\cr 
& \times  \left(\text{Im}\left\lbrace A(k_1 \tau)^2\right\rbrace\text{Im}\left\lbrace A(k_2 \tau')^2\right\rbrace\text{Im}\left\lbrace A(q_3 \tau'')B(q_1 \tau'')B(q_2 \tau'')\right\rbrace(-\tau)^{\alpha_-}(-\tau')^{\alpha_-}\right.\cr
& \left. + \ \text{Im}\left\lbrace A(k_1 \tau)^2\right\rbrace\text{Im}\left\lbrace A(q_3 \tau')B(q_1 \tau')\right\rbrace\text{Im}\left\lbrace A(k_2 \tau'')^2B^*(q_2 \tau')\right\rbrace(-\tau)^{\alpha_-}(-\tau'')^{\alpha_-}\right.\cr 
& \left. + \ \text{Im}\left\lbrace A(q_3 \tau)\right\rbrace\text{Im}\left\lbrace A(k_1 \tau')^2B^*(q_1 \tau)\right\rbrace\text{Im}\left\lbrace A(k_2 \tau'')^2B^*(q_2 \tau)\right\rbrace(-\tau')^{\alpha_-}(-\tau'')^{\alpha_-}\right.\cr
& + 1\longleftrightarrow 2 \big).
\end{align}
where $N_\zeta^{(5)}$ is defined in an analogous way to $N_\zeta^{(4)}$, and
\begin{align}
A(x)\equiv \sum_{i}\pi^{(i)}(0)s^{(i)*}(x)\ \ \ \ \ \ \ B(x)\equiv \sum_{i}b_-^{(i)}s^{(i)*}(x).
\end{align}
We also compute the contribution due to the diagram in panel $g$ of Fig.~\ref{fig:feynman 3 pt}, $N_{\zeta,g}^{(5)}$:
\begin{align}
N_{\zeta,g}^{(5)}&({\bf k}_1,{\bf q}_1-{\bf k}_1,{\bf k}_2,{\bf q}_2-{\bf k}_2,{\bf q}_3)
 =-\left(\frac{H^2}{\dot \phi_0}\right)^2\left(\frac{V'''}{H}\right)^3\frac{16\sum_{i,j} a_0^{(i)}b_-^{(i)*}|b_-^{(j)}|^2}{k_1^9 k_2^6  q_2^3 q_3^3}(q_1 q_2^2)^{\alpha_-}\cr 
&\int_{-\infty}^0\frac{d\tau}{\tau^{4}}\int_{-\infty}^0 \frac{d\tau'}{\tau'^{4}}\int_{-\infty}^{\tau'}\frac{d\tau''}{\tau''^4}(-\tau\tau'\tau'')^{\alpha_-}\cr
& \left(\text{Im}\left\lbrace(A(k_1 \tau)^2\right\rbrace\text{Im}\left\lbrace(A(k_2 \tau')\right\rbrace\text{Im}\left\lbrace(A(k_1 \tau'')\sum_{i}s^{(i)}(k_1 \tau')s^{(i)*}(k_1 \tau'')\right\rbrace\right.\cr 
&\ \ \ \ \ \ \ \ \ +1\longleftrightarrow 2\Bigg).
\end{align}

Two diagrams also contribute to the six-point function. These diagrams are shown in panels $h$ and $i$ of Fig.~\ref{fig:feynman 3 pt}.  For panel $h$, we find
\begin{align}
N_{\zeta,h}^{(6)}&({\bf k}_1,{\bf q}_1-{\bf k}_1,{\bf k}_2,{\bf q}_2-{\bf k}_2,{\bf k}_3,{\bf q}_3-{\bf k}_3)=\left(\frac{H^2}{\dot \phi_0}\right)^6\left(\frac{V'''}{H}\right)^4\frac{64 (|b_-^{(i)}|^2)^2q^3(q_1^2 q_2^2)^{\alpha_-}}{k_1^6k_2^6k_3^6q_1^3q_2^3q_3^3}\cr 
&\int_{-\infty}^0\frac{d\tau}{\tau^4}\int_{-\infty}^0 \frac{d\tau'}{\tau'^4}\int_{-\infty}^{0} \frac{d\tau''}{\tau''^4}\int_{-\infty}^{\tau''}\frac{d\tau'''}{\tau'''^4}(\tau \tau' \tau''\tau''')^{\alpha_-}\cr 
&\left(\text{Im}\left\lbrace A(k_1 \tau)^2\right\rbrace\text{Im}\left\lbrace A(k_2 \tau')^2\right\rbrace\text{Im}\left\lbrace A(k_3\tau'')\right\rbrace\text{Im}\left\lbrace A(k_3 \tau''')s^{(i)}(k_3 \tau'')s^{(i)*}(k_3 \tau''')\right\rbrace\right.\cr
&+\text{~cyc.~perm(1, 2, 3)}\big).
\end{align}
For panel $i$, on the other hand,
\begin{align}\label{eq:sixA}
N_{\zeta i}^{(6)}&({\bf k}_1,{\bf q}_1-{\bf k}_1,{\bf k}_2,{\bf q}_2-{\bf k}_2,{\bf k}_3,{\bf q}_3-{\bf k}_3)=\left(\frac{H^2}{\dot \phi_0}\right)^6\left(\frac{V'''}{H}\right)^4\frac{16 q^3(q_1 q_2 q_3)^{\alpha_-}}{k_1^6k_2^6k_3^6q_1^3q_2^3q_3^3}\cr 
&\int_{-\infty}^0\frac{d\tau}{\tau^4}\int_{-\infty}^\tau \frac{d\tau'}{\tau'^4}\int_{-\infty}^{\tau'} \frac{d\tau''}{\tau''^4}\int_{-\infty}^{\tau''}\frac{d\tau'''}{\tau'''^4}\cr 
&\left(\text{Im}\left\lbrace A(k_1 \tau)^2\right\rbrace\text{Im}\left\lbrace A(k_2 \tau')^2\right\rbrace\text{Im}\left\lbrace A(k_3\tau'')^2\right\rbrace\text{Im}\left\lbrace B(q_1 \tau''')B(q_2 \tau''')B(q_3 \tau''')\right\rbrace(-\tau \tau' \tau'')^{\alpha_-}\right.\cr 
&+\text{Im}\left\lbrace A(k_1 \tau)^2\right\rbrace\text{Im}\left\lbrace A(k_2\tau')^2\right\rbrace\text{Im}\left\lbrace B(q_1 \tau'')B(q_2 \tau'')\right\rbrace\text{Im}\left\lbrace A(k_3 \tau''')^2B^*(q_3 \tau'')\right\rbrace(-\tau \tau' \tau''')^{\alpha_-}\cr 
&+\text{Im}\left\lbrace A(k_1 \tau)^2\right\rbrace\text{Im}\left\lbrace B(q_1 \tau'')\right\rbrace\text{Im}\left\lbrace A(k_2\tau'')^2B^*(q_2\tau')\right\rbrace\text{Im}\left\lbrace A(k_3 \tau''')^2B^*(q_3 \tau')\right\rbrace(-\tau \tau' \tau''')^{\alpha_-}\cr 
&+\text{all perm(1, 2, 3)}\big).
\end{align}

As with the four-point function of $\zeta$, our task of evaluating these integrals is simplified by the fact that these integrals are IR dominated.  We keep terms leading in $\alpha_-$ and $q/k$.  As mentioned in section ~\ref{Four Point Function label}, we take $(k_i/k_j)^{\alpha_-}\simeq 1$ and $(q_i/q_j)^{\alpha_-}\simeq 1$, but not $(q_i/k_j)^{\alpha_-}$.  With this assumption, we can integrate the above expressions in a way similar to our integration of the four-point function's nested integrals.

\end{document}